\documentclass[preprint2,numberedappendix]{emulateapj}

\usepackage{siriodef}
\usepackage{capt-of}
\usepackage{graphicx}
\usepackage{subfigure}
\usepackage{verbatim}
\usepackage{microtype}

\shorttitle{Spectroscopy of Quiescent Galaxies at $1.5 < z < 2.5$ - I. Dynamics}
\shortauthors{Belli, Newman and Ellis}
\submitted{Accepted for publication in ApJ}

\begin{document}

\title{MOSFIRE Spectroscopy of Quiescent Galaxies at $1.5 < \lowercase{z} < 2.5$. \\
I - Evolution of Structural and Dynamical Properties}

\author{Sirio Belli\altaffilmark{1}, Andrew B. Newman\altaffilmark{2}, Richard S. Ellis\altaffilmark{3,4,5}}
\altaffiltext{1}{Max-Planck-Institut f\"ur Extraterrestrische Physik (MPE), Giessenbachstr. 1, D-85748 Garching, Germany}
\altaffiltext{2}{The Observatories of the Carnegie Institution for Science, 813 Santa Barbara St., Pasadena, CA 91101, USA}
\altaffiltext{3}{European Southern Observatory, Karl-Schwarzschild Str. 2, D-85748 Garching, Germany}
\altaffiltext{4}{Department of Physics and Astronomy, University College London, Gower Place, London WC1E 6BT, UK}
\altaffiltext{5}{Department of Astronomy, California Institute of Technology, MS 249-17, Pasadena, CA 91125, USA}


\begin{abstract}
We present deep near-infrared spectra for a sample of 24 quiescent galaxies in the redshift range $1.5 < z < 2.5$ obtained with the MOSFIRE spectrograph at the W. M. Keck Observatory. In conjunction with a similar dataset we obtained in the range $1 < z < 1.5$ with the LRIS spectrograph, we analyze the kinematic and structural properties for 80 quiescent galaxies, the largest homogeneously-selected sample to date spanning 3 Gyr of early cosmic history. Analysis of our Keck spectra together with measurements derived from associated HST images reveals increasingly larger stellar velocity dispersions and smaller sizes to redshifts beyond $z\sim2$. By classifying our sample according to \Sersic\ indices, we find that among disk-like systems the flatter ones show a higher dynamical to stellar mass ratio compared to their rounder counterparts which we interpret as evidence for a significant contribution of rotational motion. For this subset of disk-like systems, we estimate that $V/\sigma$, the ratio of the circular velocity to the intrinsic velocity dispersion, is a factor of two larger than for present-day disky quiescent galaxies. We use the velocity dispersion measurements also to explore the redshift evolution of the dynamical to stellar mass ratio, and to measure for the first time the physical size growth rate of individual systems over two distinct redshift ranges, finding a faster evolution at earlier times. We discuss the physical origin of this time-dependent growth in size in the context of the associated reduction of the systematic rotation.
\end{abstract}



\section{Introduction}
\label{sec:introduction}

Simulations of galaxy formation in the context of the $\Lambda$CDM cosmological model show that structure formation follows a hierarchical assembly \citep[e.g.,][]{springel05millennium}. However, in the last decade or so, this picture was initially challenged by the discovery of a population of high-redshift massive galaxies \citep{franx03, daddi04massive}. A large fraction of this population consists of quiescent objects \citep{cimatti04, daddi05}, which were formed at even earlier times and subsequently quenched. These massive quiescent galaxies are also, surprisingly, much more compact than local galaxies of similar stellar mass \citep{trujillo06, cimatti08, vandokkum08, szomoru12}. Significant observational and theoretical efforts are being directed toward improving our understanding of their evolution. We can broadly divide the life of a typical massive compact galaxy into three phases: initial star-formation, quenching, and passive evolution. Therefore the questions that need to be addressed are:
\begin{enumerate}
\item How did these massive and compact objects form, and what are their star-forming progenitors?
\item What physical processes drive the quenching of star formation?
\item What mechanisms govern their subsequent evolution?
\end{enumerate}

One of the most direct ways to explore the physical properties of galaxies is by observing their spectra. The strong absorption features found in the rest-frame optical spectra of quiescent galaxies allow us to measure their stellar velocity dispersions, which feature in important scaling relations of spheroidal systems such as the fundamental plane \citep{djorgovski87, dressler87} and the mass plane \citep{bolton07}. Velocity dispersions also allow us to estimate dynamical masses, which can be used to constrain crucial properties such as the initial mass function (IMF) and the dark matter fraction, which are otherwise difficult to measure. Even in the local universe, these key ingredients are currently poorly understood.

Besides entering the scaling relations, velocity dispersions are tightly connected to the stellar populations of quiescent systems \citep[e.g.,][]{franx08,wake12}, and are thought to be one of their most fundamental observable properties, remaining nearly unchanged during merger episodes \citep[e.g.,][]{hopkins09scalingrel, oser12}. Therefore, spectroscopic observations can be used to link high-redshift progenitors with their local descendants: by comparing galaxies at different redshifts that have the same velocity dispersion, the physical size growth of individual systems can be directly measured.

Another important aspect that can be probed by the kinematics is the presence of rotation. Since their discovery, it was noticed that high-redshift quiescent galaxies have lower \Sersic\ indices \citep{toft07, vandokkum08} and flatter shapes \citep{vanderwel11, chang13} than early-type galaxies at $z\sim0$, suggesting that these systems are more similar to local disks. Due to the difficulty of spatially resolving absorption lines, so far only one direct measurement of rotation for a quiescent galaxy beyond $z\sim1$ has been obtained \citep{newman15}.

However, at high redshift it is significantly more difficult to obtain spectroscopic data of adequate quality, partly because the spectral region of interest (the rest-frame optical) is redshifted into the observed red and near-infrared bands. Strong, variable sky emission and absorption, together with limited detector sensitivity, have made it challenging to obtain the high signal-to-noise ratio required to measure absorption features for targets at $z>1$. For this reason, most studies of high-redshift quiescent galaxies are based on only broad- or medium-band photometry, which is sufficient primarily for measuring stellar masses.

Recent progress in detector technology is finally allowing us to obtain high-quality spectroscopic data at high redshift. The upgrade of the red-sensitive detector on the LRIS spectrograph at Keck allowed us to collect the largest number of spectra to date with clearly detected absorption lines for objects at $1<z<1.5$. Our analysis of both the kinematics and the stellar populations in this sample led us to conclude that the observed size evolution of quiescent galaxies is due in part to physical growth via minor merging, and in part to the quenching of larger star-forming systems \citep{newman10, belli14lris, belli15}.

The most massive objects, however, were formed at $z>1.5$, where the size evolution has been claimed to be more rapid \citep{newman12, vanderwel14}, and spectroscopic observations are particularly difficult. A new generation of near-infrared instruments has only recently allowed the possibility of studying rest-frame optical spectra at these early epochs \citep{onodera12, onodera14, whitaker13, perez-gonzalez13, bedregal13, krogager14, newman14, mendel15}. However, only a handful of absorption-line spectra are currently available for individual galaxies in this redshift range \citep{vandokkum09, kriek09, toft12, vandesande13, hill16, barro16mosfire}. A large, homogeneous sample is needed for a robust measurement of the dynamics and stellar populations of these galaxies.

Taking advantage of the high sensitivity of the multi-object near-infrared spectrograph MOSFIRE at Keck \citep{mclean12}, we have obtained a large sample of absorption line spectra at $1.5 < z < 2.5$. We presented data for a modest initial sample in \citet{belli14mosfire}, where we also performed a preliminary analysis of the kinematics, finding that quiescent galaxies at $z\sim2$ follow the expected trends of larger velocity dispersions and smaller sizes found at lower redshifts.

In this paper we present the final sample of MOSFIRE spectra and explore the structural and dynamical properties in detail. By using the same methods developed for our previous $1 < z < 1.5$ campaign, we can ensure that our combined sample of LRIS and MOSFIRE observations is analyzed in a consistent manner. This allows us to explore trends over a wider redshift range within our combined sample, ensuring a more robust analysis than is possible by comparing high-redshift data with observations obtained in the local universe --- comparisons that often involve large systematic uncertainties.

While the MOSFIRE spectra contain a wealth of information on the stellar population properties, the present analysis focuses only on the kinematics of quiescent galaxies, which are used to explore mainly the passive phase of the evolution. In a companion paper (S. Belli et al., in preparation) we will use the same data to derive star formation histories, addressing the complementary issues of the formation and quenching of massive galaxies.

This paper is organized as follows. In Section \ref{sec:data} we present the sample selection and describe the spectroscopic observations and data reduction. In Section \ref{sec:physical_properties} we explain the methods used in deriving the physical properties of the sample, which are then presented in Section \ref{sec:hiz_properties}. We use the dynamical measurements to constrain the amount of rotation in Section \ref{sec:rotation}. In Section \ref{sec:redshift_evolution} we reconstruct the redshift evolution of the quiescent population, and we track their size growth under the assumption of fixed velocity dispersion. Finally, we summarize and discuss the main results in Section \ref{sec:conclusions}.

Throughout the paper we use AB magnitudes and assume a $\Lambda$CDM cosmology with $\Omega_M$=0.3, $\Omega_{\Lambda}$=0.7 and $H_0$= 70 km s$^{-1}$ Mpc$^{-1}$.


\section{Data}
\label{sec:data}


\subsection{Target Selection and Ancillary Data}
\label{sec:data_selection}

In order to take full advantage of the deep MOSFIRE spectra, we chose to target galaxies in fields for which abundant public data are available. In particular, we require the presence of deep high-resolution imaging from which structural parameters can be robustly measured even for $z \sim 2$ compact galaxies, and the availability of photometric data that cover a wide wavelength range. We therefore decided to carry out our observations in the fields covered by the Cosmic Assembly Near-IR Deep Extragalactic Survey \citep[CANDELS,][]{grogin11, koekemoer11}. In each of the five CANDELS fields, deep \HST\ $F160W$ observations are available, together with a rich set of photometric data. We make use of the photometric catalog assembled by the 3D-HST team \citep[version 4.1,][]{brammer12, skelton14, momcheva16}, which includes derived properties such as stellar mass and photometric redshift. We complemented this catalog with the public X-ray observations from the Chandra COSMOS Survey \citep{elvis09}, Subaru/XMM-Newton Deep Survey \citep{ueda08}, and the AEGIS-X survey \citep{laird09}.

Such an extensive data set, besides being used in our scientific analysis, allows a very efficient selection of the spectroscopic targets, which is critically important for deep observations of faint objects. We assigned a weight to each target in the public catalog according to its likelihood of yielding a detection of one or more rest-frame absorption lines. Since the observations were carried out in the $Y$ and $J$ bands, we repeated the procedure two times, one per band. For each band, the weight of an object was calculated using a combination of several factors:
\begin{itemize}
\item \emph{Photometric redshift}: highest priority was given to redshift values that would result in an ideal visibility of the main absorption features within the observed wavelength range. The ideal ranges are $1.5 < z < 1.8$ for the $Y$ band and $2 < z < 2.4$ for the $J$ band. In order to account for the uncertainty in the photometric redshifts, we gave intermediate priority to those targets with a redshift in the vicinity of the ideal range.
\item \emph{Observed near-infrared magnitude}: larger weights were given to brighter objects, for which the observations are more likely to succeed.
\item \emph{Rest-frame $U-V$ and $V-J$ colors}: our previous spectroscopic survey \citep{belli14lris, belli15} showed that this is a very robust method to identify quiescent galaxies. Objects closer to the center of the red sequence were given the highest priority.
\end{itemize}
These criteria ensure that the top priority objects are massive, quiescent galaxies at $1.5 < z < 2.5$.

Finally, we identified the regions within the CANDELS fields where the largest number of targets with high priority are located, and designed the MOSFIRE slitmasks. This introduces a potential bias towards high density environments, particularly at $z>2$, where good targets are rare. We discuss further this effect in Section \ref{sec:data_sample}.

\begin{deluxetable}{lcccl}
\tabletypesize{\footnotesize}
\tablewidth{0pc}
\tablecaption{MOSFIRE Observations \label{tab:mosfire_masks}}
\tablehead{
\colhead{Slitmask} & \colhead{Band} & \colhead{Seeing\tablenotemark{a}} & \colhead{Exp. Time} & \colhead{Run\tablenotemark{b}} \\
     &     & \colhead{(arcsec)}      & \colhead{(min)}       }
\startdata
COSMOS2					& $J$				    & 0.7	        & 484	& -\tablenotemark{c}       \\
						& $K$					& 0.8			& 60	& C \vspace{2mm} \\
COSMOS3                 & $J$    			    & 0.7	        & 534	& A       \\
						& $K$					& 0.8			& 60	& C \vspace{2mm} \\
COSMOS6	                & $Y$    		 	    & 0.7			& 312	& B		\\
		                & $H$					& 0.7		    & 84	& B \vspace{2mm} \\		
UDS1                    & $Y$   			    & 0.7		    & 330	& A       \\
		                & $H$   			    & 0.8	        & 90	& A \vspace{2mm} \\
UDS2					& $Y$					& 0.9			& 288	& C		\\
						& $H$					& 1.1			& 60	& C \vspace{2mm} \\
EGS2	                & $Y$					& 0.7			& 240	& B		\\
		                & $H$					& 0.6		    & 152	& B
\enddata
\tablenotetext{a}{The seeing is calculated from the trace of a star in the slitmask}
\tablenotetext{b}{The observing runs took place in 2014 November 25-28 (A), 2015 April 12-15 (B), and 2015 November 2-5 (C).}
\tablenotetext{c}{Data from \citet{belli14mosfire}\vspace{2mm}}
\end{deluxetable}


\subsection{MOSFIRE Spectroscopy}
\label{sec:data_spectra}

We obtained spectroscopic observations using Keck MOSFIRE during three observing runs of four nights each in 2014 and 2015. We observed a total of five slitmasks in three of the CANDELS fields (UDS, COSMOS, and EGS), and we used only data obtained in good conditions (clear sky or thin clouds, and good seeing). The exposure times varied between four and nine hours per mask. We adopted a two-point dithering pattern, with exposure times for individual frames between 120 and 180 s, and used a $0\farcs7$ slit width which yields a spectral resolution $R\sim 3500$. We observed one mask in $J$ and four masks in $Y$. For each mask we also obtained shallower $H$ or $K$ band observations with the goal of measuring \Halpha\ emission lines. Table \ref{tab:mosfire_masks} lists the details of the observations for each slitmask. We also include in this analysis the $J$-band COSMOS2 mask observed in our pilot run and presented in \citet{belli14mosfire}\footnote{In the present paper we adopt the identification numbers from the 3D-HST catalog, therefore the targets presented in \citet{belli14mosfire} have different IDs: 31719, 31769, 5517, 1966, 4126 are called in the present work, respectively, 11494, 12020, 7884, 1769, 5681.}. In order to ensure consistency, we reduced and calibrated the raw data together with the new observations, using updated procedures.

The data were reduced using the Data Reduction Pipeline\footnote{https://github.com/Mosfire-DataReductionPipeline} (DRP), which performs flat fielding, sky subtraction, cosmic ray removal, wavelength calibration, and outputs the rectified 2D spectra. From these we optimally extracted the 1D spectra \citep{horne86}, adopting the light profile (i.e., the 2D flux integrated along the wavelength direction) as the weight for each target.

Although the near-infrared sky presents strong OH lines, the AB dithering pattern, together with the optimal sky subtraction \citep{kelson03} performed by the pipeline, allows one to accurately remove the sky emission. However, the atmosphere also introduces strong absorption features. We account for these telluric features by combining our observations of A0V standard stars with atmospheric transmission models. This procedure yields not only a telluric correction but also an accurate estimate of the uncertainty introduced by the variation of the atmospheric conditions. We discuss the telluric correction in detail in Appendix \ref{appendix:telluric}.

The telluric correction also accounts for the relative flux calibration, but not for the absolute flux calibration since slit loss, air mass, seeing, and transmission typically vary between the observation of the standard star and the observation of the science targets. We calculated the absolute calibration in the following way: in each mask we positioned one slit on a relatively bright star; we then extract its spectrum in the same way as for the science targets, and integrate over the entire bandpass to obtain a photometric measurement. The absolute calibration factor is then obtained by requiring this photometric measurement to match the one from the 3D-HST catalog.

\begin{figure}[tbp]
\centering
\includegraphics[width=0.5\textwidth]{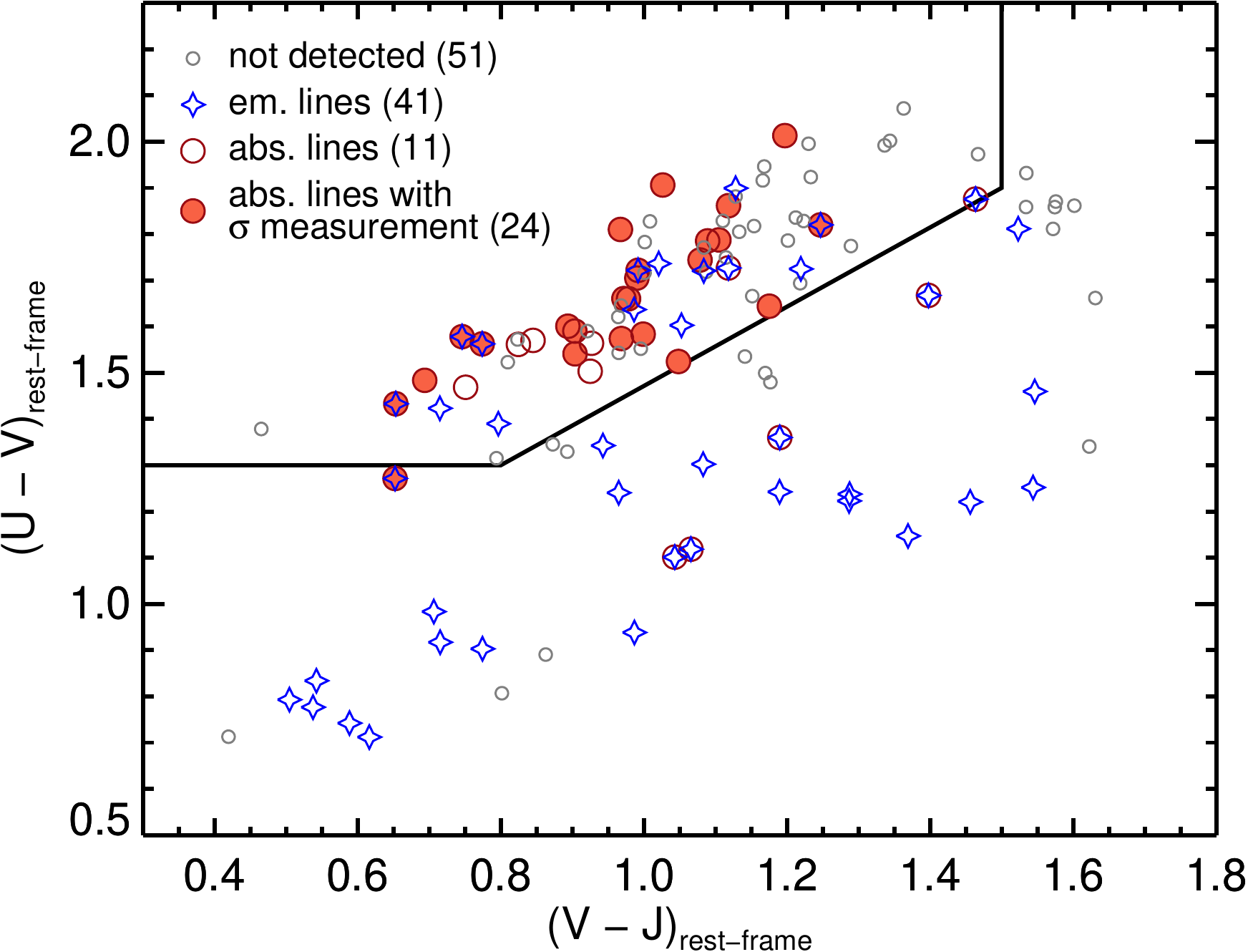}
\caption{The targeted sample on the $UVJ$ diagram. The final sample of 24 quiescent galaxies with robust velocity dispersion measurements (see Section \ref{sec:dispersions}) is shown as red filled circles. Another 11 spectra show clear absorption lines but did not yield velocity dispersion measurements (empty red circles). The 41 galaxies with detected emission lines are marked as blue diamonds (of these, 12 objects present both absorption and emission lines). Small empty circles indicate targets for which it was not possible to detect clear spectral features. The line marking the division between quiescent and star-forming galaxies is from \citet{muzzin13}.}
\label{fig:mosfire_UVJ}
\end{figure}


\begin{figure*}[tbp]
\begin{minipage}{\textwidth}
   \centering
 \raisebox{-0.5\height}{\includegraphics[width=0.45\textwidth]{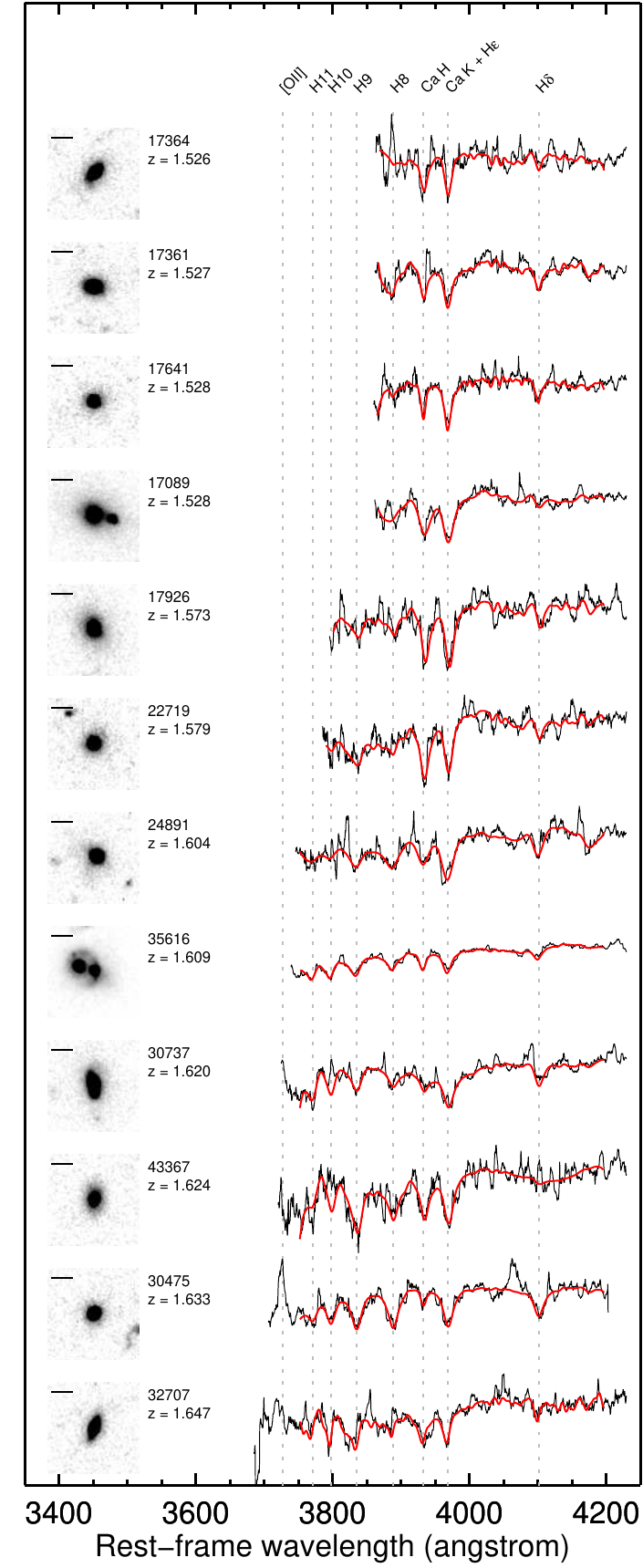}}
   \hspace*{0.01\textwidth}
 \raisebox{-0.5\height}{\includegraphics[width=0.45\textwidth]{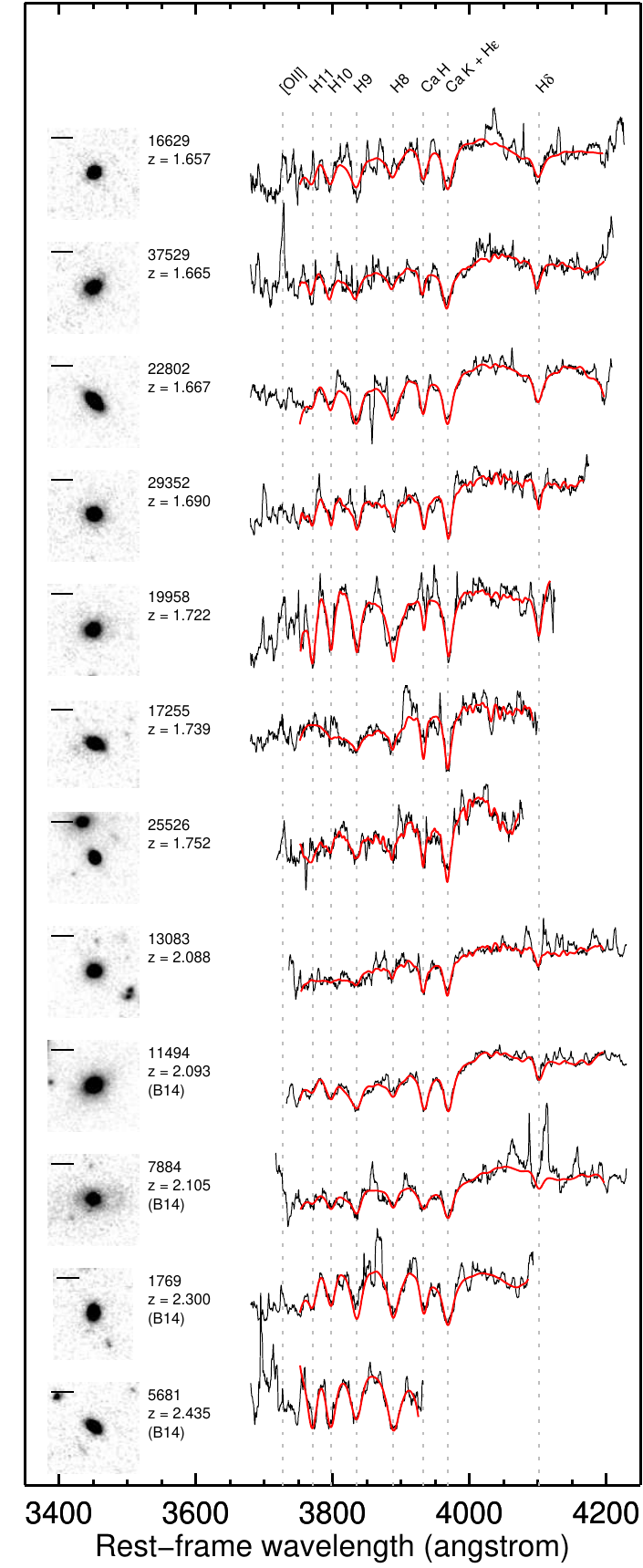}}
\end{minipage}
\caption{\HST\ images and MOSFIRE spectra for our final sample of 24 quiescent galaxies. For each object, the ID and spectroscopic redshift are indicated; the 4\arcsec cutout shows the F160W image with a 10 kpc ruler; the observed spectrum (in black) and the best-fit model (in red) are plotted. Gray dashed lines mark important spectral features. The four objects marked as B14 were already presented in \citet{belli14mosfire}.}
\label{fig:mosfire_spectra}
\end{figure*}


\begin{deluxetable*}{llccccccccc}
\tabletypesize{\footnotesize}
\tablewidth{0pc}
\tablecaption{The MOSFIRE Sample of Quiescent Galaxies \label{tab:sample}}
\tablehead{
\colhead{ID} & \colhead{Slitmask} & \colhead{R.A.} & \colhead{Decl.} & \colhead{$z$} & \colhead{\sigmae} & \colhead{$R_\mathrm{maj}$} & \colhead{n} & \colhead{q} & \colhead{$\log \Mstar/\Msun$} & \colhead{$\log \Mdyn/\Msun$}
\\
(3D-HST) & & (J2000) & (J2000) & & (km s$^{-1}$) & (kpc) & & 
} 
\startdata
17364 & COSMOS6 & 150.08154 & 2.3573 & 1.526 & $ 168 \pm 84 $ & $ 3.1 \pm 0.3 $ & 3.0 & 0.49 & $ 10.83 $ & $ 11.13 \pm 0.44 $  \\
17361 & COSMOS6 & 150.07079 & 2.3569 & 1.527 & $ 169 \pm 43 $\tablenotemark{d} & $ 2.0 \pm 0.2 $ & 2.4 & 0.66 & $ 10.80 $ & $ 10.96 \pm 0.22 $  \\
17641 & COSMOS6 & 150.09665 & 2.3597 & 1.528 & $ 142 \pm 54 $ & $ 1.4 \pm 0.1 $ & 6.0 & 0.85 & $ 10.65 $ & $ 10.50 \pm 0.33 $  \\
17089 & COSMOS6 & 150.07941 & 2.3581 & 1.528 & $ 348 \pm 57 $ & $ 5.9 \pm 0.6 $ & 4.7 & 0.88 & $ 11.56 $ & $ 11.96 \pm 0.15 $  \\
17926 & EGS2 & 215.14639 & 53.0273 & 1.573 & $ 231 \pm 39 $ & $ 5.6 \pm 0.6 $ & 4.3 & 0.71 & $ 11.01 $ & $ 11.60 \pm 0.15 $  \\
22719 & EGS2 & 215.09816 & 53.0134 & 1.579 & $ 262 \pm 51 $ & $ 2.2 \pm 0.2 $ & 6.1 & 0.92 & $ 11.03 $ & $ 11.21 \pm 0.17 $  \\
24891 & UDS1 & 34.44676 & -5.1940 & 1.604 & $ 391 \pm 71 $ & $ 2.0 \pm 0.2 $ & 2.7 & 0.86 & $ 10.85 $ & $ 11.68 \pm 0.16 $  \\
35616\tablenotemark{a} & UDS1 & 34.43073 & -5.1578 & 1.609 & $ 198 \pm 49 $ & $ 4.5 \pm 0.5 $ & 6.3 & 0.65 & $ 11.11 $ & $ 11.28 \pm 0.22 $  \\
30737\tablenotemark{a} & UDS2 & 34.58788 & -5.1758 & 1.620 & $ 307 \pm 82 $ & $ 3.9 \pm 0.4 $ & 2.2 & 0.42 & $ 11.23 $ & $ 11.79 \pm 0.24 $  \\
43367 & UDS2 & 34.56343 & -5.1271 & 1.624 & $ 299 \pm 74 $ & $ 2.7 \pm 0.3 $ & 5.2 & 0.54 & $ 11.07 $ & $ 11.46 \pm 0.22 $  \\
30475\tablenotemark{a} & UDS2 & 34.54455 & -5.1749 & 1.633 & $ 296 \pm 109 $ & $ 1.0 \pm 0.1 $ & 2.9 & 0.73 & $ 10.74 $ & $ 11.14 \pm 0.32 $  \\
32707 & UDS2 & 34.57337 & -5.1678 & 1.647 & $ 174 \pm 30 $ & $ 1.7 \pm 0.2 $ & 3.8 & 0.24 & $ 11.14 $ & $ 10.87 \pm 0.16 $  \\
16629 & COSMOS6 & 150.15811 & 2.3493 & 1.657 & $ 358 \pm 76 $ & $ 0.8 \pm 0.08 $ & 2.4 & 0.66 & $ 10.61 $ & $ 11.22 \pm 0.19 $  \\
37529 & UDS1 & 34.49199 & -5.1505 & 1.665 & $ 232 \pm 60 $ & $ 2.2 \pm 0.2 $ & 3.5 & 0.64 & $ 11.00 $ & $ 11.23 \pm 0.23 $  \\
22802 & UDS1 & 34.44692 & -5.2007 & 1.667 & $ 291 \pm 31 $ & $ 1.5 \pm 0.2 $ & 2.3 & 0.34 & $ 10.92 $ & $ 11.33 \pm 0.10 $  \\
29352 & UDS1 & 34.46959 & -5.1786 & 1.690 & $ 146 \pm 31 $ & $ 1.1 \pm 0.1 $ & 4.3 & 0.77 & $ 10.84 $ & $ 10.48 \pm 0.19 $  \\
19958 & COSMOS6 & 150.15019 & 2.3829 & 1.722 & $ 169 \pm 87 $ & $ 3.0 \pm 0.3 $ & 3.4 & 0.81 & $ 10.72 $ & $ 11.10 \pm 0.45 $  \\
17255 & COSMOS6 & 150.13335 & 2.3556 & 1.739 & $ 147 \pm 40 $\tablenotemark{d} & $ 1.9 \pm 0.2 $ & 3.6 & 0.50 & $ 10.84 $ & $ 10.76 \pm 0.24 $  \\
25526 & EGS2 & 215.10115 & 53.0270 & 1.752 & $ 134 \pm 36 $\tablenotemark{d} & $ 0.9 \pm 0.09 $ & 2.2 & 0.52 & $ 10.73 $ & $ 10.40 \pm 0.24 $  \\
13083 & COSMOS3 & 150.09610 & 2.3135 & 2.088 & $ 197 \pm 52 $ & $ 1.5 \pm 0.2 $ & 3.1 & 0.82 & $ 11.11 $ & $ 10.95 \pm 0.23 $  \\
11494\tablenotemark{b} & COSMOS2 & 150.07393 & 2.2980 & 2.093 & $ 319 \pm 26 $ & $ 3.1 \pm 0.3 $ & 4.9 & 0.79 & $ 11.67 $ & $ 11.59 \pm 0.08 $  \\
7884\tablenotemark{b} & COSMOS2 & 150.06562 & 2.2611 & 2.105 & $ 430 \pm 69 $ & $ 5.1 \pm 1.5 $\tablenotemark{e}  & 8.0 & 0.67 & $ 11.47 $ & $ 11.91 \pm 0.19 $  \\
1769\tablenotemark{b} & COSMOS2 & 150.05489 & 2.1982 & 2.300 & $ 338 \pm 46 $ & $ 1.2 \pm 0.1 $ & 2.6 & 0.71 & $ 11.17 $ & $ 11.33 \pm 0.13 $  \\
5681\tablenotemark{b} & COSMOS2 & 150.05579 & 2.2361 & 2.435 & $ 452 \pm 130 $ & $ 1.9 \pm 0.2 $ & 1.4 & 0.45 & $ 10.96 $ & $ 11.84 \pm 0.25 $ \\
\hline \\
33527 & UDS2 & 34.57123 & -5.1646 & 1.645 & \nodata & $ 2.3 \pm 0.2 $ & 2.4 & 0.90 & $ 11.03 $ & \nodata  \\
17858 & UDS1 & 34.44619 & -5.2181 & 1.663 & \nodata & $ 4.0 \pm 0.4 $ & 4.3 & 0.64 & $ 11.30 $ & \nodata  \\
24945 & UDS1 & 34.41200 & -5.1936 & 1.686 & \nodata & $ 3.8 \pm 0.4 $ & 0.5 & 0.77 & $ 10.58 $ & \nodata  \\
26536 & EGS2 & 215.09615 & 53.0278 & 1.739 & \nodata & $ 1.1 \pm 0.1 $ & 5.4 & 0.78 & $ 10.63 $ & \nodata  \\
22905 & EGS2 & 215.11676 & 53.0269 & 1.741 & \nodata & $ 4.9 \pm 0.5 $ & 8.0 & 0.93 & $ 10.80 $ & \nodata  \\
35111 & UDS1 & 34.45360 & -5.1589 & 1.822 & \nodata & $ 0.8 \pm 0.08 $ & 2.5 & 0.31 & $ 10.86 $ & \nodata  \\
9227 & COSMOS2 & 150.06176 & 2.2737 & 1.862 & \nodata & $ 1.2 \pm 0.1 $ & 3.0 & 0.65 & $ 10.96 $ & \nodata  \\
12020\tablenotemark{a,b} & COSMOS2 & 150.07460 & 2.3020 & 2.095 & \nodata & $ 2.6 \pm 0.3 $ & 5.3 & 0.54 & $ 11.27 $ & \nodata  \\
19680 & COSMOS3 & 150.07527 & 2.3794 & 2.164 & \nodata & $ 2.4 \pm 0.2 $ & 0.7 & 0.84 & $ 10.76 $ & \nodata  \\
4732\tablenotemark{b,c} & COSMOS2 & 150.05246 & 2.2455 & 2.439 & \nodata & \nodata & \nodata & \nodata & \nodata & \nodata  \\
12995 & COSMOS3 & 150.09969 & 2.3118 & 2.444 & \nodata & $ 1.4 \pm 0.1 $ & 2.4 & 0.75 & $ 11.02 $ & \nodata 
\enddata
\tablecomments{The present analysis is based only on the upper 24 objects, for which the velocity dispersion measurements are robust, as discussed in Section \ref{sec:dispersions}.}
\tablenotetext{a}{Detected in the X-ray}
\tablenotetext{b}{Presented in \citet{belli14mosfire}}
\tablenotetext{c}{Outside the CANDELS field; the ID is from \citet{belli14mosfire}}
\tablenotetext{d}{Measured using templates from the Indo-US library}
\tablenotetext{e}{Size measured from the curve of growth}
\end{deluxetable*}


\subsection{The MOSFIRE Sample}
\label{sec:data_sample}

Our MOSFIRE observations targeted a total of 115 galaxies with photometric redshift $1.4 < z < 2.5$. In Figure \ref{fig:mosfire_UVJ} we show the target sample on the $UVJ$ diagram, i.e., the rest-frame $U-V$ vs $V-J$ color-color plot, which is very effective in distinguishing between star-forming and quiescent galaxies \citep[e.g.,][]{wuyts07, williams09}. The rest-frame colors are taken from the 3D-HST catalog, and were derived by fitting models to the observed photometry. For 64 objects we identify at least one emission or absorption feature (shown respectively via blue and red symbols in Figure \ref{fig:mosfire_UVJ}) that allows us to derive a spectroscopic redshift. Of these, 35 have clear absorption lines. For 11 of these galaxies (shown as red empty circles) we could not obtain a robust spectral fit, as described in Section \ref{sec:dispersions}. Our final sample of quiescent galaxies consists of the remaining 24 galaxies for which the stellar velocity dispersion can be reliably measured (red filled circles). Remarkably, almost all the successful targets are found in the $UVJ$ selection box defined by \citet{muzzin13} to identify quiescent galaxies, confirming the effectiveness of our selection methods. 

We present the absorption line spectra for the final sample in Figure \ref{fig:mosfire_spectra}. For each target, the MOSFIRE spectrum (in black) and the best-fit model (in red, see Section \ref{sec:dispersions}) are plotted, and the $HST$ image in the F160W filter is shown. Of the 24 galaxies, 19 are observed in the $Y$ band ($1.5 < z < 1.9$), and 5 in the $J$ band ($2 < z < 2.5$), 4 of which have already been presented in \citet{belli14mosfire}. For each of the galaxies in the final sample we detect many absorption lines that are a combination of Balmer and/or CaII (H and K) lines. Other detected features include the $G$ band and the \OII\ emission line. The remaining MOSFIRE spectra in the $H$ and $K$ bands will be presented and analysed in the companion paper (S. Belli et al., in preparation).

Coordinates and spectroscopic redshifts for all the galaxies with a robust detection of absorption lines are listed in Table \ref{tab:sample}. The first 24 constitute the final sample with high signal-to-noise ratio and reliable velocity dispersion measurements. Given the scarcity of absorption line measurements at $z>1.5$, we also include in the table the spectroscopic redshifts of the other 11 galaxies with lesser quality spectra, which can be useful to test photometric redshifts or for follow-up studies.

Our spectroscopic observations confirm the high quality of the photometric redshifts. The discrepancy between spectroscopic and photometric redshifts $(z_\mathrm{phot}-z_\mathrm{spec})/z_\mathrm{spec} $ has a standard deviation of $0.03$ for absorption line systems and $0.07$ for galaxies with emission lines. Interestingly, significant outliers are only present among emission line galaxies (four objects with discrepancy larger than 0.1), while none are found among galaxies with absorption lines. The success rate, defined as the fraction of targeted galaxies for which a spectroscopic redshift was obtained via identification of absorption and/or emission lines, presents a strong dependence on the galaxy $H$ band magnitude, declining from 83\% for the brightest quartile to 43\% for the faintest one. This suggests that the number of targets missed because of large photometric redshift uncertainties is small, and that low signal-to-noise ratio is the main responsible for the lack of spectroscopic features in most cases.

Figure \ref{fig:mosfire_UVJ} shows that our final sample probes the full extent of the red sequence, except for the very red end, where many objects did not yield a detection. In Appendix \ref{appendix:samplebias} we assess this effect by comparing our MOSFIRE sample with a large, photometric sample at the same redshift drawn from the 3D-HST catalog. From the distribution of observed properties we conclude that the MOSFIRE sample is slightly biased and misses the red and faint tail of the distribution. There is no bias, however, toward more compact objects.

In order to increase the observational efficiency, we observed the areas within the CANDELS fields with the highest density of targets. In particular, some of the observed galaxies belong to the protoclusters at $z\sim2.1$ in COSMOS \citep{spitler12} and at $z\sim1.6$ in the UDS field \citep{papovich10}. High-density environments might therefore be overrepresented in the MOSFIRE sample. However, this should not affect the results of our dynamical and morphological analysis, since the difference in the size distribution of cluster and field quiescent galaxies at these redshifts appears to be negligible \citep[e.g.,][]{newman14}. This is consistent with the fact that the MOSFIRE sample has a similar size distribution as the field population, as shown in Appendix \ref{appendix:samplebias}.


\begin{figure}[htbp]
\centering
\includegraphics[width=0.5\textwidth]{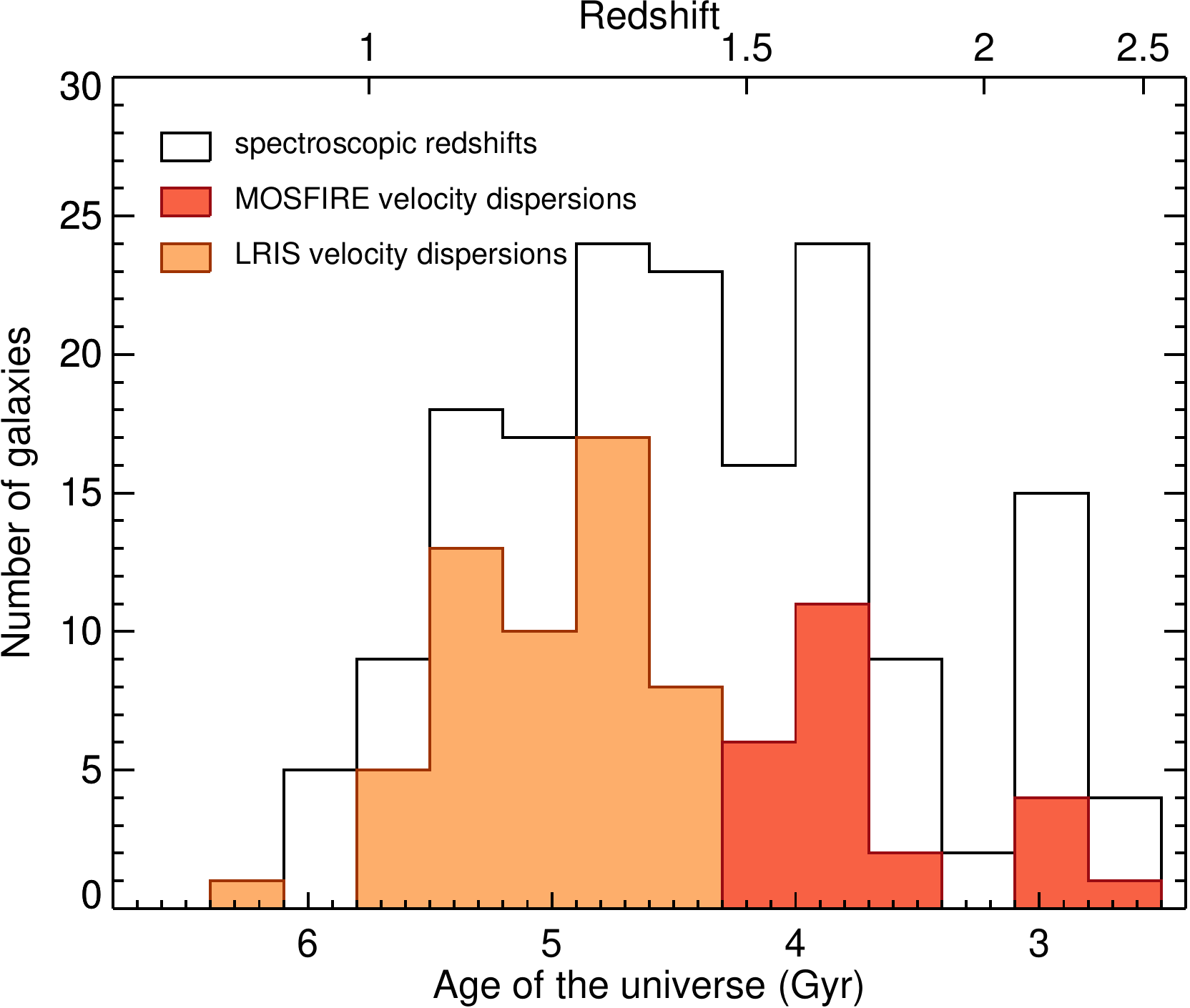}
\caption{Redshift distribution for the combined Keck sample. The black line shows the distribution of spectroscopic redshifts, including both absorption and emission line detections. The filled histograms indicate the samples of MOSFIRE (red) and LRIS (orange) spectra with a reliable measurement of velocity dispersion from absorption lines. \vspace{3mm}}
\label{fig:z_histogram}
\end{figure}

\subsection{The Combined Keck Sample}
\label{sec:keck_sample}

The MOSFIRE observations discussed in this paper represent the natural extension to higher redshift of our previous LRIS campaign \citep{newman10, belli14lris, belli15}. The two samples probe the same rest-frame wavelength range with a similar spectral resolution, and target objects selected with similar criteria from the same CANDELS fields. Furthermore, the derivation of physical properties such as sizes and velocity dispersions for the MOSFIRE sample (presented in Section \ref{sec:physical_properties}) follows the methods adopted for the LRIS sample.

The homogeneity in both data and analysis allows us to combine the LRIS and MOSFIRE samples. Figure \ref{fig:z_histogram} shows the redshift distribution of the combined sample: the total number of spectroscopic redshift measurements is 167 (black histogram). Our analysis is based on the spectra with clear absorption lines that yielded a robust measurement of the velocity dispersion. These are 56 spectra from the LRIS sample (orange histogram) and 24 from the MOSFIRE sample (red histogram), for a total of 80 galaxies. This combined Keck sample is the largest sample of deep absorption-line spectra of quiescent galaxies at $z>1$, and represents the majority of the published high-redshift stellar velocity dispersions to date. Its redshift range spans about 3 Gyr of cosmic history, and allows us to constrain the evolution of galaxy properties within a set of homogeneous high-redshift observations. This is a complementary approach to the comparison of high-redshift galaxies to the local population, which probes a larger interval in cosmic time but is potentially more affected by systematic effects.


\subsection{The Local Sample}
\label{sec:data_localsample}

In order to compare the physical properties of high-redshift galaxies with those of the local population, we select a sample of low-redshift galaxies from the Sloan Digital Sky Survey (SDSS) DR7 \citep{abazajian09}. We make use of three different catalogs: 
\begin{itemize}
\item The NYU Value Added Catalog \citep{blanton05} provides the basic properties such as multi-band photometry from the SDSS and from the Two Micron All Sky Survey (2MASS), spectroscopic redshifts, and velocity dispersions derived from absorption lines. For each object we calculate the velocity dispersion within one effective size \sigmae\ by applying the small aperture correction derived by \citet{cappellari06} to the observed velocity dispersion $\sigma$.
\item The MPA-JHU catalog \citep{kauffmann03} includes star formation rates and stellar masses derived via SED fitting assuming a \citet{chabrier03} IMF.
\item The UPenn SDSS PhotDec catalog \citep{meert15} consists of structural measurements obtained from the $r$ band imaging using \textsc{galfit} \citep{peng02}. In particular, we adopt the results of the one-component \Sersic\ fit, which include effective size, total magnitude, \Sersic\ index, and axis ratio for each galaxy.
\end{itemize}

We begin by selecting all the galaxies with spectroscopic redshift in the range $0.05 < z < 0.15$. We then exclude all the objects that are not present in all three catalogs, and those without a reliable velocity dispersion, stellar mass measurement, or surface brightness fit. These cuts exclude 15\% of the initial sample.

Because we require $J$ band photometry in order to calculate the $UVJ$ rest-frame colors, we are limited by the relatively shallow 2MASS data, which means that the high-redshift tail of the SDSS sample is incomplete. On the other hand, at very low redshifts there is a non-negligible number of galaxies brighter than $r\sim14.5$, where the SDSS observations are affected by incompleteness \citep{strauss02}, and the surface brightness fitting is more problematic \citep{meert15}. We found that the redshift range $0.08 < z < 0.10$ strikes an ideal balance between these two opposite effects, since less than 10\% of massive (above $10^{10.7}\Msun$) galaxies in this subsample are not detected by 2MASS, and only 11 objects are brighter than $r=14.5$. The sample of $J$-detected galaxies with $ 0.08 < z < 0.10$ consists of 44,922 objects.

From the observed SDSS and 2MASS photometry, we calculate the rest-frame $U-V$ and $V-J$ colors using InterRest \citep{taylor09}. Although the rest-frame colors of quiescent galaxies at $z\sim0$ are different from those of the high-redshift population, the bimodality is clearly visible on the $UVJ$ diagram. Therefore we can easily select the $UVJ$ quiescent galaxies among the local population, in the same way as we have done for our high-redshift samples. The final sample of local quiescent galaxies that we will use in the following analysis consists of 27,659 objects, of which 21,859 are more massive than $10^{10.7}$\Msun.


\section{Methodology}
\label{sec:physical_properties}

The main physical properties that we will use in our analysis are the stellar masses, the effective sizes, and the velocity dispersions. We take the stellar masses from the 3D-HST catalog, and we calculate sizes and velocity dispersions following the methods outlined in \citet{belli14lris}. In this section we discuss in detail how each of the physical properties was derived.


\subsection{Stellar Masses}
\label{sec:masses}

We adopt the stellar masses released by the 3D-HST team which were calculated with a standard spectral energy distribution (SED) fitting, presented in \citet{skelton14}. In brief, stellar population templates from the \citet{bruzual03} library were fitted to the observed photometry using the FAST code \citep{kriek09}. The \citet{chabrier03} IMF with solar metallicity, an exponentially declining star formation history, and the \citet{calzetti00} dust attenuation law were assumed.

The large number of photometric data points probing a wide wavelength range available for the CANDELS fields ensures a reliable SED fitting, which has been shown to be a robust method for measuring stellar masses \citep[e.g.][]{wuyts09,muzzin09sed}. The formal uncertainties are typically small, especially when a large number of accurate photometric points is used. However, systematic uncertainties due to the choice of star formation history, dust attenuation law, and spectral templates typically dominate the error budget, and are difficult to estimate \citep[for a review see][]{conroy13review}. For simplicity, we assume a stellar mass uncertainty of 0.2 dex for every target in our sample, which is representative of the uncertainty due to the use of different methods and templates \citep{mobasher15} but does not include the effect of changing the IMF.

In order to ensure consistent size and mass measurements, for each galaxy we scale the 3D-HST stellar mass and total F160W flux to match the flux measured by our surface brightness fit (described in Section \ref{sec:sizes}). We also apply a distance modulus correction, to account for the discrepancy between the photometric redshifts used in the SED fits and our accurate spectroscopic measurements. Both corrections are small, with mean values of 0.03 dex and -0.02 dex respectively. The corrected stellar masses for the MOSFIRE sample are listed in Table \ref{tab:sample}.

We note that the stellar masses for the LRIS sample were calculated in \citet{belli14lris} via SED fitting using various sources of public photometry. To ensure consistency among our combined Keck sample, we compare our mass measurements for the LRIS galaxies with the values in the 3D-HST catalog, which are available for 54 out of 56 objects. We find no systematic difference between the two measurements, with a mean offset of 0.02 dex and a standard deviation of 0.1 dex.


\subsection{Structural Properties}
\label{sec:sizes}

The sizes and other basic structural properties were derived for each galaxy using the public \HST\ data in the F160W band, which corresponds to the rest-frame optical emission. We use \textsc{galfit} \citep{peng02} to fit a two-dimensional \Sersic\ profile to the observed surface brightness for each object. Despite the small sizes of high-redshift galaxies, simulations show that this technique provides reliable measurements \citep{mosleh13, davari14}. Neighboring objects are identified using SExtractor \citep{bertin96} and are either masked out or fit simultaneously, according to their distance and brightness. We fix the background level following \citet{newman12}, and combine isolated bright stars to derive the point-spread function (PSF).

The fitting procedure outputs a number of parameters for each galaxy, of which the most physically interesting are the \Sersic\ index $n$, the axis ratio $q$, and the half-light semi-major axis \Rmaj, which we list in Table \ref{tab:sample}. We assume an uncertainty of 10\% on the size measurement, which has been shown to be a good approximation to the true error \citep{vanderwel08, newman12}. This is confirmed by a comparison of our measurements with those derived by \citet{vanderwel12} on the same data set: the discrepancy in size is only 3\% on average, with a standard deviation of 7\%.

Only one of the objects (7884) could not be fit with a single \Sersic\ profile, probably because of the bright asymmetric halo visible in the F160W image. As discussed in \citet{belli14mosfire}, we measure the size of this object via analysis of the curve of growth, obtaining a value that is in agreement with the result of the \Sersic\ fit within 30\%.


\subsection{Velocity Dispersions}
\label{sec:dispersions}

Velocity dispersions were measured by fitting broadened templates to the observed MOSFIRE spectra using the Penalized Pixel-Fitting routine (pPXF) of \citet{cappellari04}. The instrumental resolution was measured for each object by fitting a Gaussian profile to the sky emission lines. The measured dispersion was corrected for both the instrumental resolution ($40-50$ km s$^{-1}$) and the resolution of the template spectra (between 50 and 100 km s$^{-1}$). Only the wavelength region $3750\AA < \lambda < 4200\AA$ was considered for the fit (ensuring we exclude the emission line \OII), but the exact wavelength range used varied for each object depending on redshift and physical position on the slitmask. Within the range used, the spectral pixels that are significantly contaminated by sky emission were masked out.

We adopted the \citet{bruzual03} library of synthetic spectra of stellar populations, allowing for a combination of single bursts with different ages. The exact template used in the spectral fitting can have a large impact on the measurement of the velocity dispersion: if the line depth - which for Balmer lines is a measure of stellar age - is not correctly fit, then the line width measurement will also be biased. Furthermore, the rotational velocity of a star depends strongly on its spectral type, with the largest values for A stars \citep[up to hundreds of km s$^{-1}$; e.g.,][]{nielsen13}. We test for potential systematics by repeating the spectral fit using different templates, taken from the Indo-US library of observed stellar spectra \citep{valdes04}, first using only spectra of F and G stars, then including also spectra of A stars. In most cases the velocity dispersions do not change significantly when including A stars. This is not surprising, because the pPXF algorithm automatically selects the templates that best represent the observed spectrum and, since the MOSFIRE spectra contain age-sensitive features, the fit is able to determine that A stars do not constitute the majority of the stellar population. However, there are a few galaxies for which the velocity dispersion is very sensitive to the inclusion of A-type stellar templates, and we exclude them from our sample. The only objects with high signal-to-noise ratio spectra that do not pass this test are 4732 and 24945, which are in the star-forming region of the $UVJ$ diagram. For these galaxies we do expect an important contribution to the absorption lines from A stars. This confirms that our test is able to determine the stellar population composition, and that the exact intrinsic broadening of the template spectra does not represent an issue for our sample.

We also tested for the effect of changing the wavelength range, the masking of the sky emission, and the degree of the additive and multiplicative polynomials used in the spectral fit. For a detailed discussion of the fit and the estimate of the uncertainty on the velocity dispersion, which is due in roughly equal parts to random errors and systematic effects (including template mismatch), we refer the reader to \citet{belli14lris}. The tests reveal that, out of the 35 spectra with clear detection of absorption features, the velocity dispersion measurements for 11 galaxies are not robust, and we discard them.

We note that three objects in our final sample (indicated in Table \ref{tab:sample}) are not fully resolved when using the \citet{bruzual03} library, which has a resolution of $\sim100$ km s$^{-1}$. For these galaxies we adopt instead the velocity dispersions obtained with the Indo-US library, which has a resolution of $\sim30$ km s$^{-1}$. This is unlikely to introduce any bias in the sample, since the agreement between the results obtained with the two libraries for the 21 objects that are well resolved is very good: the discrepancy in the measurement of \sigmae\ has a mean of 2\% and a standard deviation of 14\%. However, these low values of velocity dispersion are close to the intrinsic width of some of the stellar spectra, even if A stars are excluded from the template library (F0 stars can have rotational velocities up to about 100 km s$^{-1}$). For these objects the velocity dispersion measurements are therefore more sensitive to the exact choice of the template.

For the final sample of 24 galaxies, we apply a 5\% aperture correction \citep{vandesande13,belli14lris} to the measured velocity dispersions, thus obtaining \sigmae, the velocity dispersions within the effective size, which are shown in Table \ref{tab:sample}. The average relative uncertainty is 25\%. Four objects were already presented in \citet{belli14mosfire}: their spectra have been newly reduced and the velocity dispersions re-measured. The new values agree within the uncertainties with the previous ones for all galaxies except one (5681). For this object we adopt the new measurement, which has a more conservative error bar following more extensive testing.


\subsection{AGN emission}
\label{sec:agn}

We assess the importance of active galactic nucleus (AGN) emission for the MOSFIRE sample using two independent methods. First, we examine X-ray detections in the CANDELS fields and identify 4 galaxies in our sample, which we report in Table \ref{tab:sample}. Second, we calculate the IRAC colors and find that only one object in our sample (12020, also detected in X-rays) fulfills the criterion given by \citet{donley12} to identify AGN emission. We therefore conclude that the majority of the sample is not affected by strong AGN activity.


\begin{figure*}[htbp]
\centering
\includegraphics[width=0.9\textwidth]{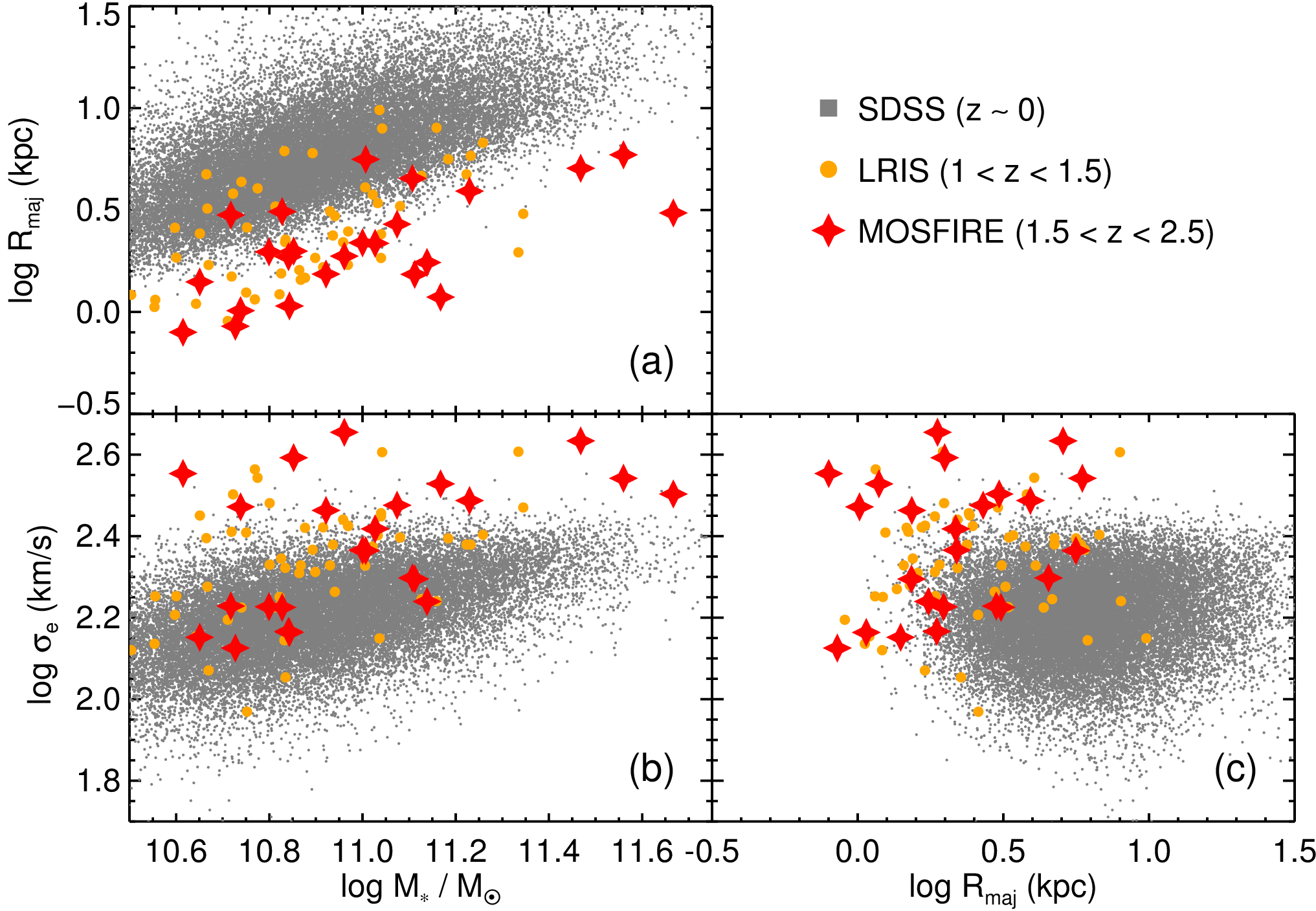}
\caption{Distribution of $UVJ$-quiescent galaxies in the 3-dimensional parameter space of size, velocity dispersion, and stellar mass. Three samples at different redshifts are shown: local galaxies from the SDSS survey (small gray points), galaxies at $1<z<1.5$ observed with Keck LRIS (orange points), and galaxies at $1.5<z<2.5$ observed with Keck MOSFIRE (red stars).}
\label{fig:cube_plot}
\end{figure*}

\section{Physical Properties of High-Redshift Quiescent Galaxies}
\label{sec:hiz_properties}


\subsection{Stellar Masses, Sizes, and Velocity Dispersions}

In Figure \ref{fig:cube_plot} we show the distribution of the three key physical properties for our Keck sample at high redshift and for the local quiescent galaxies. The distributions of stellar masses, effective sizes, and velocity dispersions are clearly different for galaxies at $z>1$ and galaxies at $z\sim0$. High-redshift galaxies have smaller sizes and larger velocity dispersions compared to local objects with the same stellar mass.

Comparing the properties of local and high-redshift galaxies is generally a challenging task because of the subtly different methods used to derive them. However, we argue that the difference in the distribution of galaxies at high and low redshift observed in Figure \ref{fig:cube_plot} is a robust result, not driven by observational uncertainties or systematic effects. First, all the objects shown have been selected as being quiescent using a consistent criterion, based on the rest-frame $UVJ$ colors. Second, the physical properties have been derived using the same methods at low and high redshift: the major axes come from \Sersic\ fits to the observed 2D surface brightness maps using GALFIT; the stellar masses from fitting templates to the broadband photometry; and the velocity dispersions from fits to the rest-frame optical spectra. One caveat is that for high-redshift galaxies the Balmer absorption lines play a significant role in the determination of the velocity dispersion, which is not true for local quiescent objects. However, we note that at high redshift the range of stellar ages is much less and thus it is reasonable to assume the Balmer lines are as representative kinematic tracer as metal lines. Third, incompleteness is unlikely to cause the differences shown in Figure \ref{fig:cube_plot}: in Appendix \ref{appendix:samplebias} we show that our MOSFIRE observations, despite missing some of the faintest quiescent galaxies, are unbiased in size. Also, the LRIS sample is virtually unbiased in size, brightness, and colors above $10^{10.8}$ \Msun\ \citep{belli15}.

The new MOSFIRE data show that the trend towards smaller sizes and larger velocity dispersions continues beyond $z \sim 1.5$. This result confirms, for the first time with a relatively large and homogeneous sample, the suggestions made in early studies, which observed increasingly larger velocity dispersions at higher redshift \citep{vandokkum09, cappellari09, cenarro09, toft12, onodera12, vandesande13, bezanson13}, including the preliminary results from our MOSFIRE campaign \citep{belli14mosfire}.

The local distribution on the mass-size plane is not symmetrical: while there is a significant tail of large galaxies, there seems to be a sharp edge at small radii below which only a few objects are found. This region of the diagram has been dubbed the \emph{zone of exclusion} \citep{bender92,cappellari13XX}, and constitutes an empirical upper limit to the compactness of local systems. Figure \ref{fig:cube_plot}a clearly shows that at $z>1$ a large fraction of the population lies in the zone of exclusion. This has already been shown, using a larger photometric sample, by \citet{newman12}, and suggests a physical evolution of individual systems. We can now use our spectroscopic sample to make the same argument about the distribution on the mass - velocity dispersion plane (shown in Figure \ref{fig:cube_plot}b), where the zone of exclusion translates into an upper limit in velocity dispersion. Again, we find that a significant population of high-redshift objects lies in the region avoided by local galaxies.


\subsection{Dynamical Masses}
\label{sec:dynamical_masses}

Stellar velocity dispersion and effective size, together with surface brightness, define a parameter space where local elliptical galaxies lie in a narrow \emph{fundamental plane} \citep{djorgovski87, dressler87}. The fundamental plane is also in place at high redshift \citep[e.g.,][]{treu05, vanderwel05}. However, the fact that the luminosity of a passively evolving galaxy is not constant makes it difficult to separate a possible evolution of the fundamental plane from the normal aging of stellar populations \citep[see, e.g.,][]{vandesande14}. It is therefore more useful to consider the \emph{mass plane}, which is obtained by replacing surface brightness with the more physically meaningful stellar mass \citep[e.g.,][]{bolton07,hyde09,cappellari13XV}. This relation between mass, velocity dispersion, and size can be easily interpreted as a consequence of the virial theorem which, for a pressure-supported system, predicts that the total gravitational mass must be proportional to $\sigmae^2 \R$, where \R\ is the effective size. We therefore define the \emph{dynamical} mass as:
\begin{equation}
\label{eq:mdyn}
	\Mdyn = \frac{\beta(n) \, \sigmae^2 \, \Rmaj }{G} \; .
\end{equation}
Our choice differs from the definitions adopted by other works \citep[including our previous analysis of the LRIS sample,][]{belli14lris} in two respects.

\begin{figure}[tbp]
\centering
\includegraphics[width=0.45\textwidth]{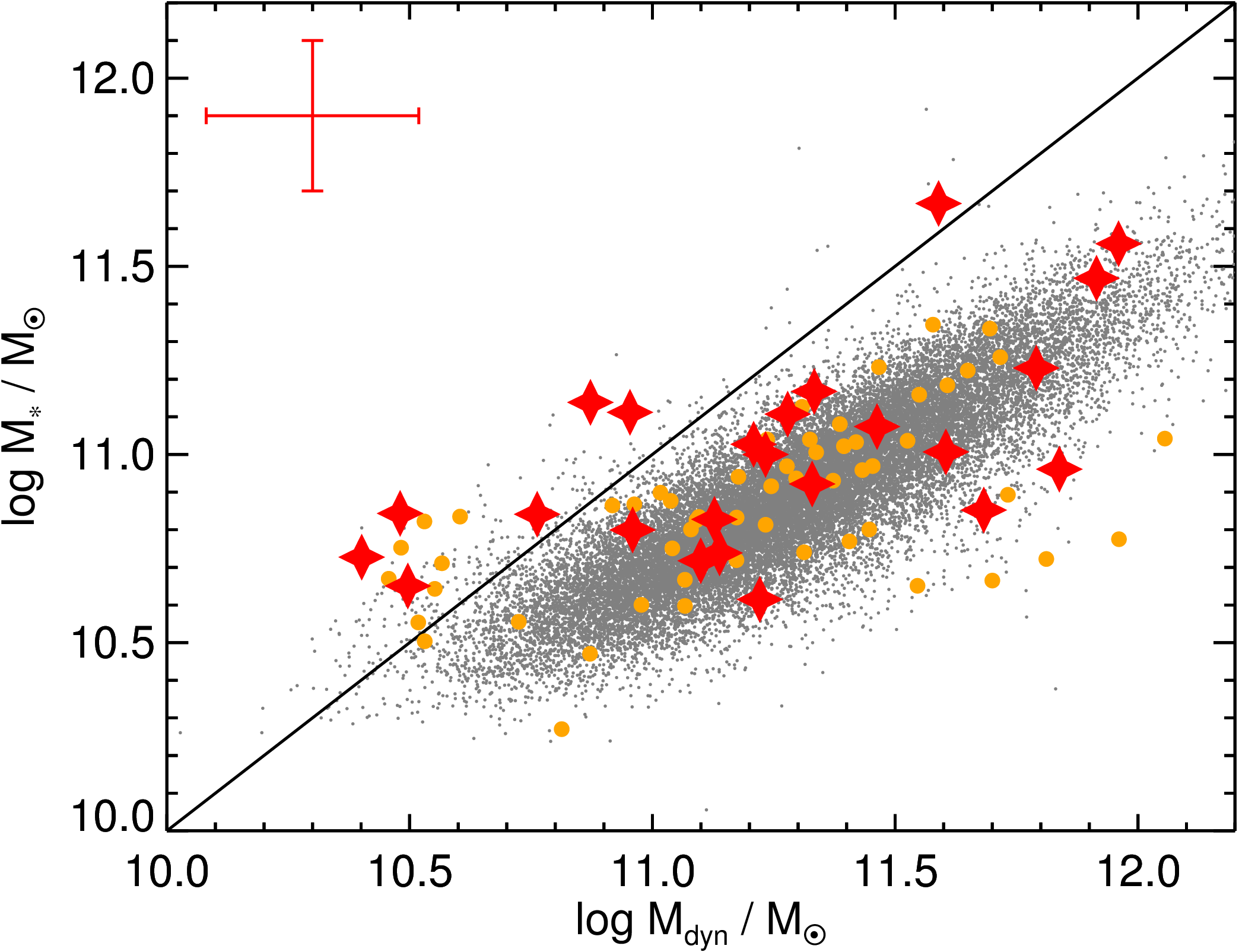}
\caption{Stellar mass versus dynamical mass (calculated using Equation \ref{eq:mdyn}) for quiescent galaxies. Symbols as in Figure \ref{fig:cube_plot}. The solid black line indicates the one-to-one relation. The median error bars for the MOSFIRE sample are shown on the top left.}
\label{fig:massplane}
\end{figure}

First, by using a proportionality constant $\beta(n)$ that is a function of the \Sersic\ index $n$, we explicitly allow for a weak homology, where galaxies with different surface brightness profiles follow slightly different scaling relations. We adopt the theoretical value for spherical isotropic models \citep{cappellari06}: $\beta(n) = 8.87 - 0.831 n + 0.0241 n^2$. Interestingly, \citet{cappellari06} found that a constant value $\beta = 5.0$ gives a smaller scatter in the local mass plane and is therefore a better choice. However, their sizes were calculated by fitting a de Vaucouleurs profile to the observed curve of growth, which differ significantly from the sizes used in this work, calculated via \Sersic\ fits to the 2D surface brightness distribution. When using a \Sersic\ profile, allowing for weak homology yields dynamical masses that are in better agreement with stellar masses \citep{taylor10} and with masses obtained via Jeans modeling \citep{cappellari13XV}.

Second, we use the major axis \Rmaj\ rather than the circularized radius $\Rmaj \sqrt{q}$. While for spheroidal objects and face-on disks the two quantities are very similar, a significant difference arises when considering inclined disks. For these galaxies the axis ratio reflects the inclination rather than the intrinsic shape, and the circularized radii can be significantly smaller than what would be measured if the systems were face-on. Major axes are therefore more robust, and have been recommended both for observational studies \citep{cappellari13XV} and when comparing to numerical simulations \citep{hopkins10}.

Using Equation \ref{eq:mdyn}, we calculate the dynamical masses for the low and high-redshift samples. In Figure \ref{fig:massplane} we compare these dynamical masses to the stellar masses obtained via SED fitting. Remarkably, and despite an increase in scatter, the MOSFIRE sample at $z>1.5$ occupies the same region in the \Mdyn-\Mstar\ plane as the local population. This extends to higher redshift the earlier result obtained with our LRIS sample \citep{belli14lris}. We also note the presence of a small number of objects in the unphysical region where $\Mstar > \Mdyn$. It is difficult to establish if these galaxies are simply scattered in that region due to the relatively large uncertainties, or if other effects must be invoked to explain the measurements, such as galaxy non-homology \citep[e.g.,][]{peraltadearriba14}.

Figure \ref{fig:massplane} represents a reformulation of the galaxy properties shown in Figure \ref{fig:cube_plot}, modulo a small correction which is function of the \Sersic\ index. The fact that on the \Mdyn-\Mstar\ diagram the observed distribution is much tighter means that this is the approximately edge-on view of the mass plane. However, this sequence is different from the virial expectation, as is clear from the fact that the galaxy distribution in Figure \ref{fig:massplane} does not lie on the one-to-one relation. Not only there is a vertical offset, but the mass plane is also \emph{tilted}. Our data suggest that the offset and tilt for the high-redshift sample are consistent with those for the local population.

If we calculate the dynamical masses using circularized radii instead of major axes, the result is qualitatively unchanged. The only differences are that the fundamental plane has a slightly shallower slope (i.e., the tilt increases), and there are less outliers with large \Mdyn\ compared to \Mstar. These outliers have elongated shapes, hence their circularized radii are significantly smaller than their major axes. The nature of these galaxies will become clear in the next Section, while in Section \ref{sec:redshift_evolution} we will explore the relation between the stellar and dynamical masses further.


\section{Rotation in Quiescent Galaxies}
\label{sec:rotation}

Beside the size growth, other important changes affect the structure of quiescent galaxies during their evolution. Morphological studies based on imaging data have shown that high-redshift quiescent systems tend to be flatter and have a lower \Sersic\ index than the local population \citep[e.g.][]{toft07,vandokkum08, vanderwel11, chang13}. These observations suggest that rotation might play an important role in $z\sim2$ quiescent systems; however, dynamical measurements are needed to directly confirm this possibility.

\subsection{The Relation between Kinematics and Structure}

\begin{figure}[tbp]
\centering
\includegraphics[width=0.45\textwidth]{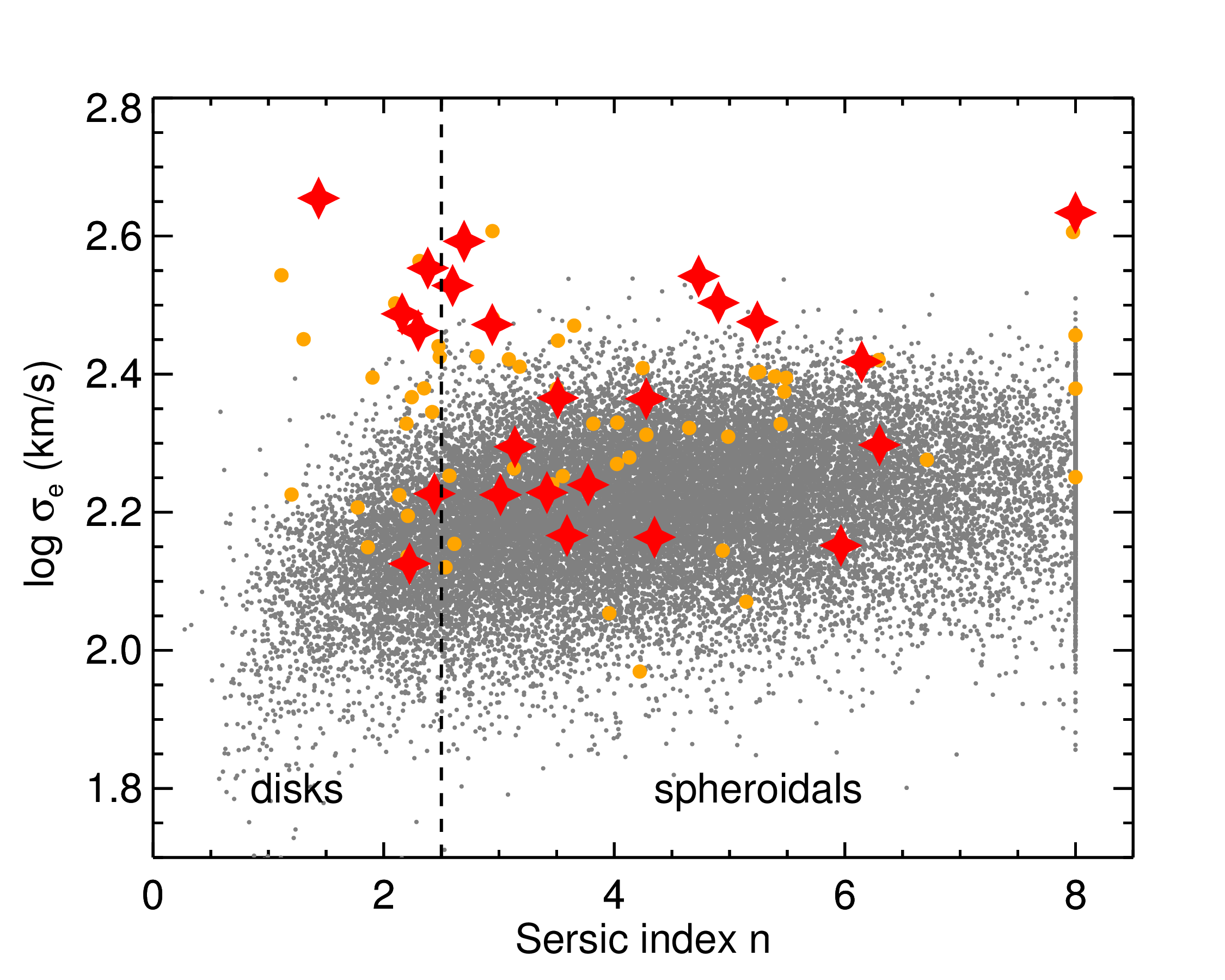}
\caption{Velocity dispersion versus \Sersic\ index $n$. Symbols as in Figure \ref{fig:cube_plot}. The vertical dashed line marks the $n=2.5$ threshold value used to separate disks from spheroidals.}
\label{fig:sigma-n}
\end{figure}
 
Previous studies have typically compared the morphological properties at low and high redshift for galaxies at a given stellar mass. Instead, we can now compare the morphology of galaxies \emph{at fixed velocity dispersion}, which is considered a more fundamental property, as we will discuss in Section \ref{sec:fixed_sigma}. We plot the relation between \sigmae\ and the \Sersic\ index $n$ in Figure \ref{fig:sigma-n}, for both the Keck sample and the local SDSS population of quiescent galaxies. The vertical dashed line marks the often adopted (yet somewhat arbitrary) $n=2.5$ threshold for identifying disky systems. We find a population of high-redshift disks with large velocity dispersions, which is completely absent from the local sample. Furthermore, while at $z\sim0$ the velocity dispersions clearly decrease toward smaller values of $n$, the trend seems to be inverted at $z\sim2$. This is a new, independent suggestion that the structure of quiescent galaxies is qualitatively different at high redshift: a simple scaling of the effective size would not be able to match the $z\sim2$ population to the local one.

The relation between structure, as measured via the \Sersic\ index, and kinematics has important consequences for our interpretation of the measured velocity dispersions. If some of the galaxies in the Keck sample are disks, then the observed value of \sigmae\ must include the contribution of rotational velocity.


\subsection{Dynamical Evidence for Rotational Support}

The definitive test for the presence of rotating disks in quiescent galaxies at $z\sim2$ would be the measurement of a rotation curve. This technique is widely used for local objects and for high-redshift galaxies with strong emission lines. However, the median size of the galaxies in our Keck sample is 2.3 kpc, and the typical seeing of $0\farcs8$ corresponds to about 6 kpc at $z\sim1-2$, thus making it impossible to spatially resolve the galaxies. In fact, spatially resolving absorption lines at high redshift has proved to be exceptionally difficult, due to the small angular sizes of typical sources and demanding requirements on the signal-to-noise ratio of the continuum. Beyond the $z\sim1$ sample of \citet{vanderwel08rotation}, the only measurement is that of \citet{newman15}, which took advantage of strong gravitational lensing to detect significant rotation in a quiescent galaxy at $z=2.6$.

\begin{figure}[tbp]
\centering
\includegraphics[width=0.5\textwidth]{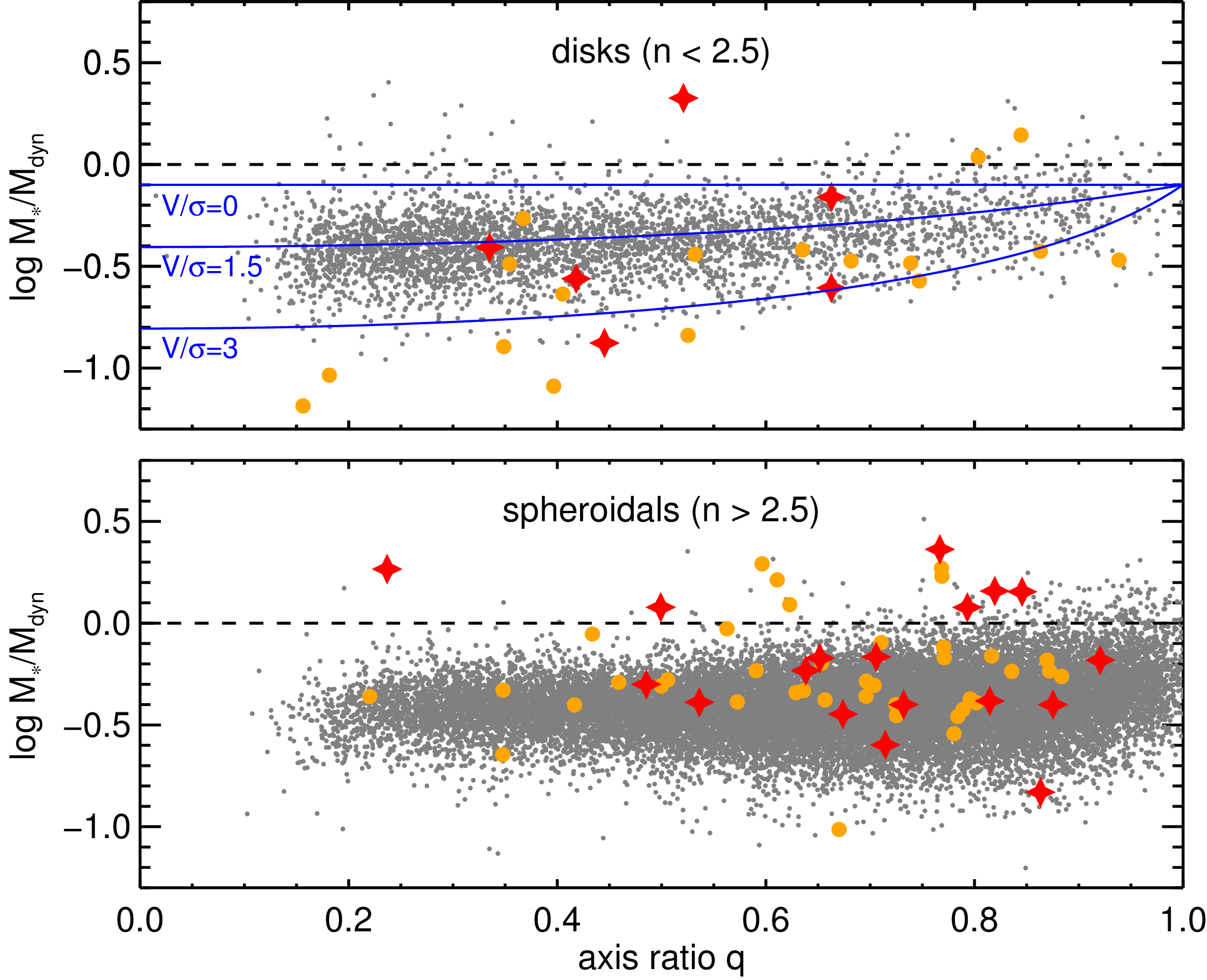}
\caption{Stellar-to-dynamical mass ratio as a function of the axis ratio $q$, for disks (\Sersic\ index $n < 2.5$, top panel) and spheroidals ($n > 2.5$, bottom panel). Symbols as in Figure \ref{fig:cube_plot}. In the top panel, the three blue curves represent the prediction for an idealized system given in Equation \ref{eq:prediction}, for three values of $V/\sigma$.}
\label{fig:rotation}
\end{figure}

However, we can indirectly probe for the presence of rotation by exploring how random and ordered motion affect the unresolved kinematics. If rotation is indeed significant in high-redshift quiescent galaxies, we expect to see a variation of the measured velocity dispersion as a function of the inclination: edge-on systems will have larger values compared to face-on systems, where only the random component contributes to the measured velocity dispersion \citep[see, e.g.,][for applications of this method to unresolved emission lines]{vandokkum15, price16}. This change in the measured velocity dispersion will cause, in turn, a corresponding change in the dynamical mass \Mdyn. Since the stellar mass does not depend in any way on the inclination, the quantity \Mstar/\Mdyn\ is ideal for such test. In the top panel of Figure \ref{fig:rotation} we plot the mass ratio \Mstar/\Mdyn\ versus the axis ratio $q$ for the disk galaxies ($n < 2.5$) at low and high redshift. While the local population has a roughly flat distribution, the Keck sample has a significantly steeper distribution, with an excess of dynamical mass at low values of $q$ and viceversa. This is qualitatively in agreement with a scenario where rotational support is significantly more important at high redshift. In the bottom panel we show the spheroidal objects (with $n > 2.5$), for which \Mstar/\Mdyn\ appears to be fairly independent of the axis ratio at both low and high redshift, as expected for pressure-supported systems.

We can calculate the expected trend of \Mstar/\Mdyn\ with axis ratio by considering a simple model of galaxy kinematics. A purely rotating, unresolved disk observed through a slit would produce an observed velocity dispersion (defined as the second moment of the velocity distribution along the line of sight) $\sigma_{obs} = \gamma \, V \, \sin i$, where $V$ is the circular velocity and $i$ is the inclination (defined so that $i=0$ for face-on systems). The conversion factor $\gamma\sim0.6-1$ has been determined observationally \citep[e.g.,][]{rix97, weiner06} and theoretically \citep[e.g.,][]{vandokkum15} for gas emission. Because the structure of gas can be different from that of the stars, we adopt the value derived from the stellar kinematics of local early-type galaxies in the ATLAS$^{3D}$ sample, for which \citet{cappellari13XV} derive $\gamma = 0.66$. We do not account for the misalignment between the slit and the kinematic major axis, since we can assume that most of the light falls within the slit independently of the orientation of the disk.

In the general case where both random and circular motions are present and not resolved, what we measure is in fact the combination of the two components:
\begin{equation}
	\sigma_{obs}^2 = \sigma^2 + \gamma^2 \, V^2 \, \sin^2 i = \sigma^2 \left( 1 + \gamma^2 \left(\frac{V}{\sigma} \right)^2 \sin^2 i \right) \;, 
\end{equation}
where $\sigma$ is the intrinsic velocity dispersion. The relative contribution of the two components to the dynamical equilibrium of a system is typically measured in terms of the ratio $V/\sigma$. We now use $\sigma_{obs}$ in place of \sigmae\ in the definition of dynamical mass, Equation \ref{eq:mdyn}, obtaining:
\begin{equation}
\label{eq:mdyn_obs}
	\Mdyn = \frac{\beta(n) \, \sigma^2 \, \Rmaj }{G} \left( 1 + \gamma^2 \left(\frac{V}{\sigma} \right)^2 \sin^2 i \right)  \; .
\end{equation}
Although the true inclination is not known, it can be estimated from the axis ratio $q$ assuming that galaxy disks are circular and have a thickness $q_z$:
\begin{equation}
\label{eq:inclination}
	\sin^2 i = \frac{1-q^2}{1-q_z^2} \; .
\end{equation}
Finally, we use this relation in Equation \ref{eq:mdyn_obs}, and normalize the dynamical mass to the value we would measure if the system were face-on:
\begin{equation}
\label{eq:prediction}
	\frac{ \Mdyn(q) }{ \Mdyn(q=1) } = 1 + \gamma^2 \left(\frac{V}{\sigma} \right)^2 \frac{1-q^2}{1-q_z^2} \; .
\end{equation}
We assume a vertical thickness $q_z = 0.2$, which is approximately the minimum axis ratio found in both the SDSS and the Keck samples \citep[see also][]{chang13}, and we plot in the top panel of Figure \ref{fig:rotation}, in blue, three curves corresponding to Equation \ref{eq:prediction} for $V/\sigma = $ 0, 1.5, and 3. The three curves have been normalized to $\log \Mstar/\Mdyn = -0.1$ at $q=1$. Clearly the high-redshift sample is well described by a larger value of $V/\sigma$ compared to the local universe. We can infer an approximate value $V/\sigma \sim 1.5$ at $z\sim0$ and $V/\sigma \sim 3$ at $z>1$. The exact value of $V/\sigma$ depends on the choice of $\gamma$, which is somewhat uncertain. Our measurement of $V/\sigma \sim 1.5$ for the local population is, in fact, slightly higher than other estimates \citep[e.g.,][]{vanderwel08rotation, emsellem11}, and suggests that our assumed value for $\gamma$ is too low. In this case the $V/\sigma$ value for the high-redshift sample of disks would also decrease. However, the ratio between the value of $V/\sigma$ for the high-redshift and local populations of quiescent disks does not depend on $\gamma$. We can then robustly conclude that for disky quiescent galaxies rotation is about twice as important at high redshift compared to the local universe. This result is based on the comparison of local and high-redshift galaxies at fixed axis ratio, and is therefore independent on the precise form of size measurement adopted for dynamical masses (i.e., circularized or major axis). It is also largely independent of the virial factor, since we only considered galaxies in the narrow range of \Sersic\ indices $1 < n < 2.5$, where $\beta(n)$ varies by only 15\%.

In summary we find important evidence that many quiescent systems at high redshift harbour stellar disks with significant rotation. Since these likely evolve into local spheroidals, a mechanism must be found to reduce the associated angular momentum. We will discuss this further in conjunction with the measured size growth rate in Section 7.


\section{Redshift Evolution of Quiescent Galaxies}
\label{sec:redshift_evolution}

In this section we analyze the evolution of masses, sizes, and velocity dispersions of quiescent galaxies, extending to higher redshift our earlier analysis of the LRIS sample \citep{belli14lris}. The main advance over our earlier work lies in having a homogeneous set of data spanning a wide redshift range, mitigating the need to rely primarily on comparisons with local samples.

\subsection{Evolution of the Stellar-to-Dynamical Mass Ratio}
\label{sec:massratio}

\begin{figure}[tbp]
\centering
\includegraphics[width=0.45\textwidth]{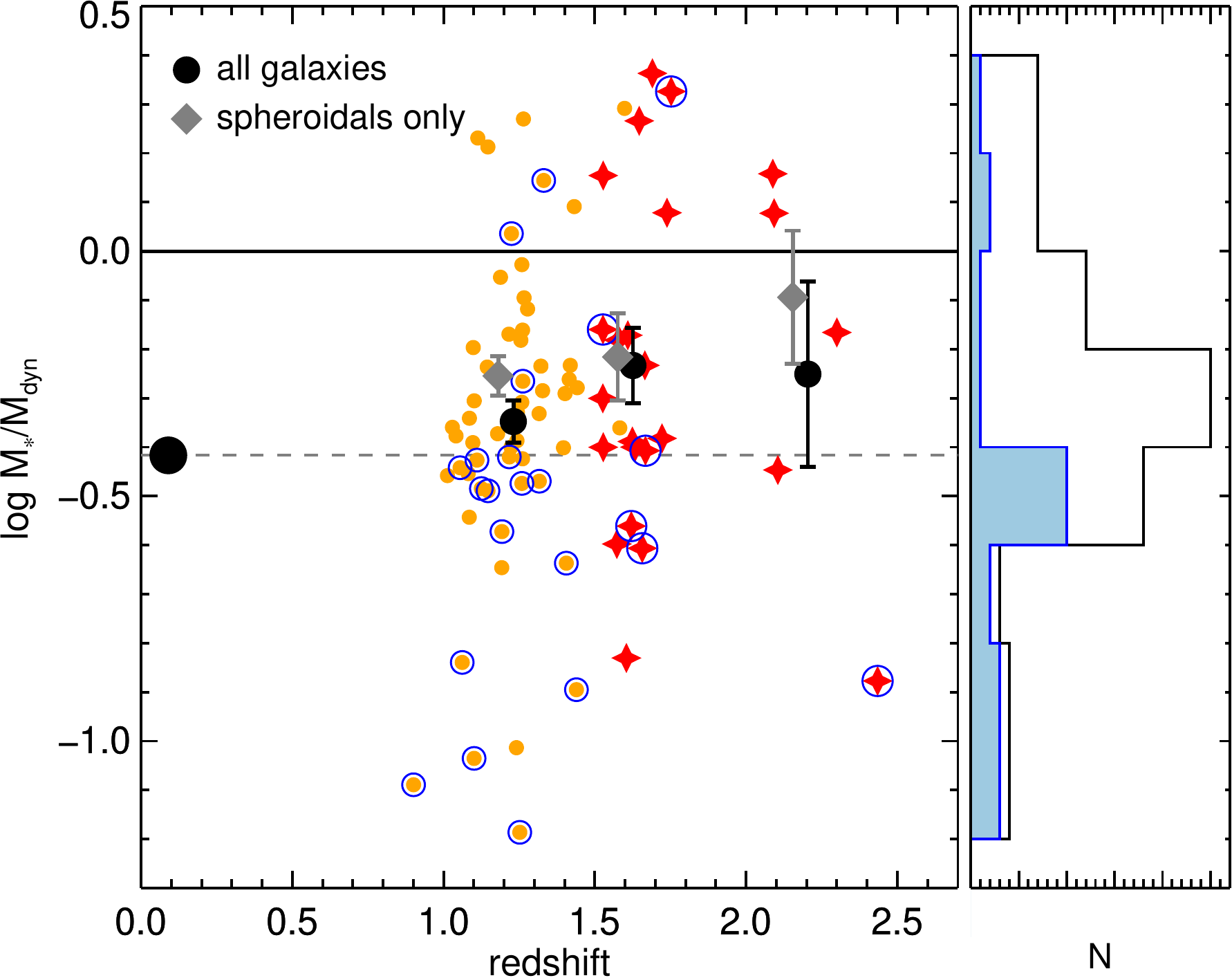}
\caption{Stellar-to-dynamical mass ratio as a function of redshift. The dashed line marks the mean value for the SDSS sample. Individual high-redshift galaxies are plotted using the same symbols as in Figure \ref{fig:cube_plot}, and objects with $n<2.5$ are identified by blue circles. For each of the three redshift bins the average value is shown in black, while gray diamonds indicate the average values calculated for spheroidal ($n>2.5$) galaxies only, with a small horizontal shift for clarity. In the right panel, the black line shows the total distribution of galaxies, while the blue filled histogram represents the distribution of disks.}
\label{fig:massratio}
\end{figure}

The stellar-to-dynamical mass ratio \Mstar/\Mdyn\ is an important observable quantity, because it is linked to fundamental properties such as dark matter fraction and IMF. In Figure \ref{fig:massratio} we plot \Mstar/\Mdyn\ as a function of redshift for the galaxies in the Keck sample, and mark the local mean value with a dashed line. As we showed in the previous section, the stellar-to-dynamical mass ratio can be strongly affected by the inclination, if significant rotation is present. Since the \Sersic\ index proved to be a good indicator of galaxy structure, we use the $n<2.5$ criterion to divide the Keck sample into disky and spheroidal galaxies. We find 23 disks in the Keck sample, corresponding to 29\% of the total, and we mark them with blue circles in Figure \ref{fig:massratio}. Although disks are only a third of the total sample, they constitute the majority of galaxies with very low values of \Mstar/\Mdyn. This is confirmed by the mass ratio distribution shown in the right panel for both the total sample (black line) and the subsample of disks (blue filled histogram). We find that 74\% of the disks lie below the local mean value, while only 16\% of the spheroidals are found in this region.

The fact that disks and spheroidals have different distributions has an important impact on the redshift evolution of the stellar-to-dynamical mass ratio. In Figure \ref{fig:massratio} we show the mean values of \Mstar/\Mdyn\ in three redshift bins, for both the total sample (black points) and the subsample of spheroidals (gray diamonds). While the total sample is marginally consistent with the local value, the spheroidal population presents a significant evolution with redshift. Combining the three redshift bins we find an evolution between the SDSS sample and the $z>1$ population of $0.10\pm0.04$ dex, significant at the 2.7-$\sigma$ level, while if we consider the spheroidals only we obtain $0.179\pm0.035$ dex, which is a 5-$\sigma$ detection. This result is qualitatively unchanged if we adopt circularized radii instead of major axes in the definition of dynamical masses (the significance becomes 2.8-$\sigma$ and 4-$\sigma$ for the full sample and the spheroidal subsample respectively).

This is the first time that such a large, homogeneous sample of dynamical masses is available at high redshift, thus allowing us to obtain statistically significant results. Among previous studies, some claimed an increase of \Mstar/\Mdyn\ at $z>1.5$, but the sample sizes were too small for conclusive results \citep{toft12, vandesande13, bezanson13massplane, belli14mosfire}. However, we suggest that the comparison between SDSS and our Keck sample could potentially be affected by systematic effects which are not captured in the formal uncertainty quoted above. Examples include subtle differences in the methods used to measure velocity dispersions and stellar masses. One way to overcome this limitation is to explore the evolution within our homogeneous high-redshift sample. Interestingly, for the spheroidal population we find tentative evidence of a redshift evolution within the Keck sample, with a stellar-to-dynamical mass ratio increasing with redshift. However, the statistical significance is too low to draw robust conclusions.

It is important to note that for the disks in our sample the dynamical masses do not take into account explicitly the contribution of rotation, and for this reason they should be considered as formal quantities rather than actual mass measurements. The evolution of the observed \Mstar/\Mdyn\ for these systems is therefore not necessarily a relevant physical property. Conversely, the dynamical masses measured for the spheroidal population should be reliable. The increase in the stellar-to-dynamical mass ratio with redshift could then be indicative of an evolution in the stellar IMF, or alternatively of the dark matter fraction. Numerical simulations show that as the effective size of quiescent galaxies increases with cosmic time, the relative contribution of dark and baryonic matter within \R\ can significantly evolve \citep[e.g.,][]{hilz13}. However, we caution against a simplistic interpretation of the observed evolution in \Mstar/\Mdyn\ because of the intrinsic difference in the structure of quiescent galaxies at high redshift and those in the local universe. Furthermore, since the mass plane is tilted (see Figure \ref{fig:massplane}), the exact value of \Mstar/\Mdyn\ depends not only on the morphology but also on the mass range probed by each sample.

\subsection{Evolution at Fixed Velocity Dispersion}
\label{sec:fixed_sigma}

One of the main challenges for observational studies of galaxy evolution is to correctly link observed objects with their progenitors at higher redshifts, and draw conclusions on the physical evolution of \emph{individual} systems, as opposed to the observed evolution of a galaxy population. Simply comparing the local quiescent population with the quiescent galaxies observed at high redshift, as done in Figure \ref{fig:cube_plot}, gives limited insight, since the composition of the population changes with time, an effect dubbed \emph{progenitor bias} \citep{vandokkum96}. In the present section we make use of a physically motivated assumption to constrain the physical evolution of individual galaxies.

Following our previous analysis of the LRIS sample presented in \citet{belli14lris}, we assume that the stellar velocity dispersion does not change significantly throughout the evolution of a quiescent galaxy, and that galaxies with large dispersions were all formed at high redshift. We justify this approximation as follows:

\begin{enumerate}
\item From a theoretical point of view, the main process that is expected to influence the structure of quiescent galaxies is galaxy merging. Numerical simulations and theoretical arguments indicate that dissipationless galaxy mergers leave the central velocity dispersions largely unchanged \citep[e.g.][]{nipoti03,hopkins09scalingrel,oser12}, despite the potentially large growth in mass.
\item Observationally, archaeological studies of local quiescent galaxies have shown that the velocity dispersion is a very good proxy for stellar age, and that at fixed dispersion there is no correlation between size and age \citep{vanderwel09,graves09_II}. A large body of evidence also suggests that the velocity dispersion is the best predictor for color, star formation activity, and mass-to-light ratio \citep[e.g.,][]{trager00,bernardi05,wake12,cappellari13XV}.
\item High-redshift observations independently support this scenario. Despite the paucity of large samples of velocity dispersion measurements at high redshift, it is possible to use the virial relation to estimate inferred velocity dispersions from the observed sizes and masses \citep{franx08}. The number density of galaxies with inferred velocity dispersion above a critical value $\sigmae \sim 280$ km/s is constant with redshift at least up to $z\sim1.5$ \citep{bezanson11}, suggesting that this property is remarkably stable for massive galaxies.
\end{enumerate}

\begin{figure}[tbp]
\centering
\includegraphics[width=0.45\textwidth]{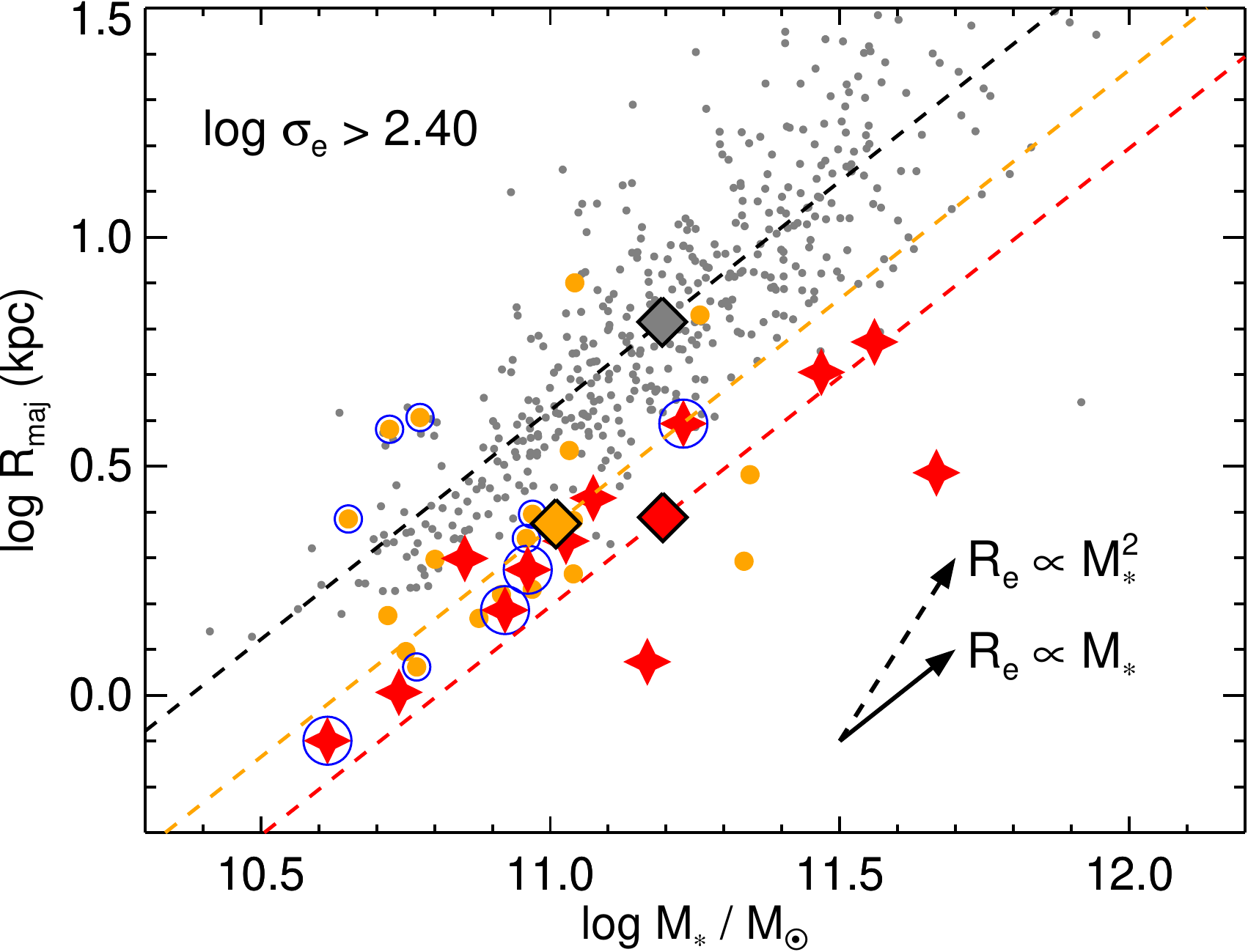}
\caption{Evolution in the mass-size plane for quiescent galaxies with $\log \sigmae > 2.40$. Symbols as in Figure \ref{fig:cube_plot}, with blue circles indicating objects with $n<2.5$. For each of the samples, the dashed line represents a linear fit with a fixed unitary slope, and the diamond marks the average value (excluding galaxies with $n<2.5$). The solid and dashed black arrows show the evolutionary tracks with slope $\alpha=1$ and 2, respectively.}
\label{fig:fixedsigma}
\end{figure}

In Figure \ref{fig:fixedsigma} we plot all the quiescent galaxies at low and high redshift with velocity dispersions larger than a threshold of $\log \sigmae = 2.40$, or $\sigmae \sim 250$ km/s, which corresponds to stellar ages older than 10 Gyr for local objects \citep{graves09_II}. Following the arguments presented above, we can consider all the objects in the figure to be connected by an evolutionary link, since their velocity dispersions do not change much with cosmic time, and at the same time there are not many new galaxies formed with such high velocity dispersions below $z\sim2$. By comparing the properties of these objects at different redshifts we can thus capture the true physical evolution, free from progenitor bias. Furthermore, by taking a large bin defined by a lower limit in velocity dispersion, we are effectively considering a more relaxed assumption of fixed \emph{ranking} in velocity dispersion (see \citealt{belli14lris} for more details).

Figure \ref{fig:fixedsigma} shows that high-redshift objects are clearly smaller and less massive than local galaxies with the same value of velocity dispersion. Here, too, we need to be careful in the interpretation of the observed velocity dispersions, which might contain a large contribution from rotational motion for the disks in the sample. We mark the galaxies with \Sersic\ index $n<2.5$ with blue circles, and we notice that these tend to be larger than the spheroidals. Interestingly, three of the four LRIS galaxies that are above the local relation happen to be disks. When removing the disks, the Keck sample is even more offset toward smaller sizes and stellar masses. 

For the first time we are now able to base our analysis on a homogeneous sample that spans a wide redshift range, from $z \sim 1$ to $z \sim 2.5$, corresponding to 3 Gyr of cosmic history. We can then measure the physical growth in three distinct redshift bins and obtain an estimate of the \emph{time derivative} of the size evolution. The average values of stellar masses and effective sizes for the three samples at $z \sim 0$, $z \sim 1.3$, and $z \sim 2$ (excluding the disks) are shown as diamonds. As the MOSFIRE sample is not exactly matched to the LRIS sample and probes a population with slightly larger stellar masses, we cannot simply connect these average values, as confirmed by the fact that the average mass is larger for the highest redshift bin, while we expect galaxies to grow in stellar mass with cosmic time. Instead, we make an additional simplification, assuming that a galaxy population of fixed velocity dispersion will have a unitary slope on the mass-size diagram. This is reasonable given that the ratio $\Mstar/\Mdyn$ is approximately constant, and at fixed $\sigmae$ this implies $\R \propto \Mstar$. A linear fit to the SDSS points yields a slope of $0.96 \pm 0.03$, thus verifying the assumption. We then fit lines of unitary slope (shown as dashed lines in Figure \ref{fig:fixedsigma}) to the three samples, and focus on the evolution of the normalization. The linear fits yield a size evolution of $0.26 \pm 0.06$ dex between $z\sim0$ and $z\sim1.3$, and a comparable value, $0.17 \pm 0.10$ dex, between $z\sim1.3$ and $z\sim2$. However, the two redshift intervals correspond to very different time intervals, and the evolution is much faster at high redshift when one compares the growth \emph{rate}: 0.13 dex/Gyr at $z\sim1.5$ compared to only 0.03 dex/Gyr between $z\sim1.3$ and $z\sim0$. This result does not change significantly when the disks are included in the calculation, in which case we obtain growth rates of 0.15 dex/Gyr and 0.03 dex/Gyr for the two redshift intervals. We note that potential systematics affecting the comparison of high-redshift and local observations can in principle affect this result. However, we stress that our measurement of the the growth rate between the LRIS and MOSFIRE samples will be unaffected by these issues.

Another constraint that can be obtained from the analysis of the mass-size diagram is the slope $\alpha$ of individual tracks. This is an important quantity, because numerical simulations can predict the values of $\alpha$ that characterize different physical processes. In Figure \ref{fig:fixedsigma} we show two examples of such tracks: $\alpha=1$ (i.e., $\R \propto \Mstar$), which is the evolution expected for major mergers, and $\alpha=2$ (i.e., $\R \propto \Mstar^2$), which represents the theoretical upper limit for the growth caused by minor mergers. While the LRIS sample suggests a value of $\alpha = 1.4 \pm 0.2$ \citep{belli14lris}, fully consistent with the expectation for minor mergers, we cannot make a similar statement for the MOSFIRE sample since it probes a different mass range, as discussed above. We note, however, that any slope below unity would not be able to explain the evolution of the linear fits shown in Figure \ref{fig:fixedsigma}. We conclude that our data suggest that $\alpha > 1$ also at $z>1.5$, which would rule out major mergers as the dominant mechanism behind the size growth at early times.


\subsection{Evolution on the Mass Plane}

If galaxies grow in size and mass, as implied by our analysis at fixed velocity dispersion, then their position on the mass plane must also change as a function of cosmic time. In Section \ref{sec:massratio} we explored the redshift evolution of the stellar-to-dynamical mass ratio, which represents the distance of a galaxy from the mass plane. Now we look directly at how the position of quiescent galaxies on the mass plane changes with cosmic time.

\begin{figure}[tbp]
\centering
\includegraphics[width=0.5\textwidth]{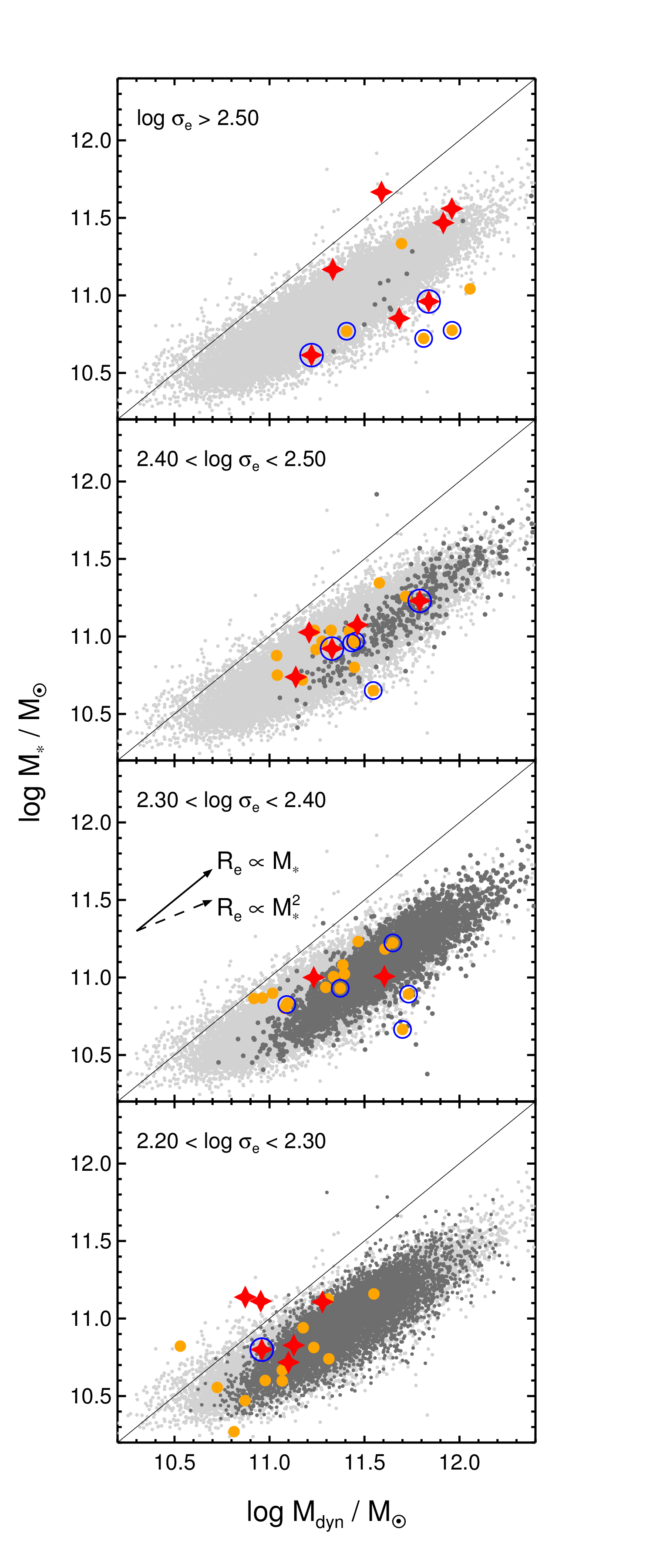}
\caption{Stellar mass versus dynamical mass for quiescent galaxies at $0 < z < 2.5$, split into bins of velocity dispersion. Symbols as in Figure \ref{fig:cube_plot}, with blue circles indicating objects with $n<2.5$. For comparison, in each panel the light gray points show the total SDSS quiescent population. As in Figure \ref{fig:fixedsigma}, the solid and dashed arrows represent the cases for $\alpha=1$ and 2, respectively, assuming evolution at constant velocity dispersion and \Sersic\ index.}
\label{fig:massplane_sigmabins}
\end{figure}

We show the mass plane again in Figure \ref{fig:massplane_sigmabins}, where we plot galaxies at low and high redshift in four bins of velocity dispersions. In the context of evolution at fixed velocity dispersion, we can consider each bin as representing the same galaxy population at different cosmic times, and interpret the differences between low and high-redshift galaxies as due to a physical evolution. Again we mark galaxies that are disks according to their \Sersic\ index, as this has important implications on their measured kinematics.

The first interesting aspect of Figure \ref{fig:massplane_sigmabins} is that the bin with the largest velocity dispersions ($\sigmae > 320 $ km/s) is mainly populated by high-redshift galaxies. This might suggest that the velocity dispersion of these very massive galaxies must decrease with cosmic time. Because these objects are frequently found in overdensities (particularly at $z>2$, most of the massive galaxies in our sample belong to a protocluster), they might be the progenitors of local brightest central galaxies (BCGs). BCGs are known to experience more major mergers than typical galaxies, due to their peculiar position at the center of the cluster \citep[e.g.,][]{delucia07}, and this might explain why they apparently experience a different evolution. However, almost half of these objects are disks, and in these cases the \emph{measured} velocity dispersion would naturally decrease if the rotational motion is transformed into random motion.

The remaining three bins of velocity dispersion show a different picture. At each value of \sigmae, high-redshift galaxies generally have lower dynamical and stellar masses than their local counterpart, consistent with merger-driven growth. Individual galaxies are therefore evolving approximately \emph{within} the mass plane, although the evolution is systematically different for spheroidals and disks, which tend to have an excess of dynamical mass compared to the local mass plane. Since $\Mdyn \propto \sigmae^2 \; \Rmaj$, the evolution on the \Mdyn-\Mstar\ plane is tightly linked to the evolution in the mass-size plane. As an illustration, in Figure \ref{fig:massplane_sigmabins} we show the two examples of growth discussed above, $\alpha=1$ and $\alpha=2$, assuming fixed velocity dispersion and \Sersic\ index.

Finally, we point out that the high-redshift points in each bin form a rather tight distribution, contrary to what one might expect given the increase in observational uncertainties with redshift. We confirm this quantitatively by measuring the standard deviation along the mass plane, i.e., the standard deviation of the quantity $(\Mstar + \Mdyn) / 2$. The dispersion for the Keck sample is smaller than that of the local population by 0.09 dex, 0.06 dex, and 0.00 dex for the three lower bins. Interestingly, \citet{wellons16} find a wide spread in the evolutionary tracks of massive compact galaxies in the Illustris simulation: from $z\sim2$ to $z\sim0$ the compact population goes from a factor of 3 dispersion in stellar mass to a factor of 20 \citep[see also][]{clauwens16}. In the simulation, the amount of growth is directly linked to the number of mergers experienced by each galaxy. The increase in the stellar mass scatter with cosmic time may represent a further independent confirmation that merging is the main driver for the mass and size growth of quiescent galaxies.


\section{Summary and Discussion}
\label{sec:conclusions}

\subsection{Summary}

We presented a sample of 24 deep rest-frame optical spectra of massive quiescent galaxies at $1.5 < z < 2.5$ obtained with Keck MOSFIRE. We combine this with our Keck LRIS spectra at $1 < z < 1.5$, thus obtaining the largest sample of absorption-line spectroscopy for quiescent galaxies at high redshift, containing 80 objects. The new, combined sample presents two important advantages over our previous analysis: it extends the redshift range thus probing earlier cosmic times, and allows for the first time a measurement of the dynamical evolution within a homogeneous high-redshift sample, mitigating the need to rely on comparisons with local data.

The three most important results of our analysis are:
\begin{enumerate}
\item We find that, among the galaxies classified as disks according to their \Sersic\ indices, edge-on galaxies have systematically larger dynamical masses than face-on systems. This is consistent with a significant contribution of rotation to the observed kinematics, which is spatially unresolved. We conclude that quiescent galaxies at high redshift are more rotationally supported than their local counterparts: $V/\sigma$ for the disks in our Keck sample is about twice that found in local quiescent disks.
\item The evolution of the ratio between stellar and dynamical mass, an important observable quantity for quiescent galaxies, depends on the morphology of the sample considered. The average value of \Mstar/\Mdyn\ at $z>1$ is marginally consistent with the local value for the overall quiescent population, but the subsample of spheroidal objects presents a statistically significant evolution of $0.18\pm0.04$ dex.
\item In order to infer the evolution of individual galaxies, we made the theoretically and observationally motivated assumption that galaxies evolve at constant velocity dispersion. We were then able to measure the size evolution in two redshift bins, and we conclude that the growth is significantly faster at earlier cosmic times: 0.13 dex/Gyr at $z\sim1.5$ compared to only 0.03 dex/Gyr over $0 < z < 1.3$.
\end{enumerate}

We caution that we rely on a comparison with local data for some of our results, particularly the evolution of the stellar-to-dynamical mass ratio. This can potentially be affected by systematic differences in the derivation of stellar masses and velocity dispersions. Such limitations are common to most high-redshift studies, and their exploration is beyond the scope of the present work. However, we stress that the discovery of significant rotational support in quiescent galaxies at high redshift, which is perhaps our most important finding, is largely independent of these effects, since it does not rely on a direct comparison of dynamical measurements obtained at different redshifts, but on a comparison of trends of dynamical quantities at fixed epochs.

\subsection{The Progenitors of Massive Quiescent Galaxies}

The formation of massive quiescent galaxies is still not well understood, but represents a critical phase in the evolution of cosmic structures. Since these are the first systems to undergo the quenching of star formation, understanding the early phase of their evolution might reveal important clues about the physical mechanisms behind galaxy quenching. One way to explore their formation is to identify the population of progenitors of these systems. Different galaxy populations have been proposed as possible progenitors of massive quiescent galaxies, such as sub-millimeter galaxies \citep[e.g.,][]{tacconi08, toft14}, or compact star-forming galaxies \citep{barro13}, which in turn can be produced by reducing the size of a larger progenitor \citep[e.g.][]{dekel14,zolotov15} or by quenching at very early times, when the universe was denser \citep[e.g.,][]{wellons15}.

Our observations offer new ways to connect massive quiescent galaxies with their potential progenitors. In particular, our detection of a significant rotational component suggests that some progenitors must be rotating disks. Compact star-forming galaxies at $z\sim2$, for example, show tentative evidence of rotation \citep{vandokkum15}. If these are credible progenitors, the quenching process must preserve some fraction of the rotational motion. Among the possible quenching mechanisms, gas-rich major mergers could be consistent with the observation of rotating remnants, as showed by numerical simulations \citep[e.g.,][]{wuyts10}.

\subsection{The Evolution of Massive Galaxies}

The internal properties of massive quiescent galaxies at $z>1$ are qualitatively different from those of their local counterparts. In addition to the detection of significant rotation, this is robustly confirmed by the fact that we found a population of $z>1$ quiescent galaxies with low \Sersic\ indices and large velocity dispersions which are not present in the local universe. The so-called passive phase of their evolution cannot therefore be truly passive, but some processes such as galaxy merging must be in place to explain the observed change in the structure of massive quiescent galaxies.

The observed size growth is another important aspect, and perhaps the most studied one, of the evolution of these massive objects. In the present study we confirm and extend to higher redshift our earlier results \citep{belli14lris}, according to which quiescent galaxies physically grow in stellar mass and particularly in size. The slope of growth on the mass-size plane is consistent, at least up to $z\sim1.3$, with the expectations for minor mergers, but not with the predictions for pure major mergers. Given the connection between the structural and dynamical properties of galaxies, it is very likely that the mechanism responsible for the size growth is also the one causing a decrease in the rotation of the disky objects. Numerical simulations have shown that minor mergers would indeed be able to cause the disruption of the disks \citep{naab14}. However, our finding of an accelerated growth rate at $z>1.3$ makes it difficult to reconcile the merger rate required to explain the size growth with the one that is observed at these redshifts \citep{newman12}. It remains to be seen whether other physical processes are partly responsible for the fast growth observed at $z\sim2$.

The fact that the size of individual galaxies increases does not rule out a substantial contribution of progenitor bias to the size evolution of the overall quiescent population \citep[e.g.,][]{carollo13}. As shown by the analysis of the stellar populations in our LRIS sample \citep{belli15}, at $z\sim1.5$ the growth of individual systems can be of the same order as the contribution of newly-quenched, larger galaxies. Given the fast rate at which the composition of the quiescent population changes at $z\sim2$, it is necessary to find a reliable way to physically connect galaxies observed at different redshifts. We made the simple assumption of an evolution at constant stellar velocity dispersion, as expected from theoretical and observational arguments. However, the discovery of significant rotation in these systems makes it difficult to interpret the observed kinematics in terms of intrinsic velocity dispersions. In the present work we have avoided this problem by excluding the disks, identified via their surface brightness profile, from our analysis. Ideally, simulations of galaxy evolution would provide also predictions for the observed unresolved velocity dispersions, which could then be easily compared to the observations.

We conclude by summarizing the scenario that is most consistent with the observational evidence presented here and in our previous works. Massive quiescent galaxies form at high redshift from star-forming progenitors, from which they retain the compact sizes and, in part, the rotational motion. The evolution of the quiescent population is then governed in part by the addition of recently quenched galaxies, which tend to be large and therefore increase the average size, and in part by a mechanism that causes individual systems to physically grow in mass and size. This mechanism, at the same time, likely destroys the rotation that was initially present, and transforms the early compact systems into the large, massive early-type galaxies that populate the local universe. \\


We thank Trevor Mendel and the VIRIAL team for many useful discussions. We also acknowledge Sadegh Khochfar, Eva Wuyts, Alan Meert and Carlo Nipoti for helpful discussions. We acknowledge Adi Zitrin for completing the observation of one of the MOSFIRE masks. RSE acknowledges support from the European Research Council through an Advanced Grant FP7/669253.
The authors recognize and acknowledge the very significant cultural role and reverence that the summit of Mauna Kea has always had within the indigenous Hawaiian community. We are most fortunate to have the opportunity to conduct observations from this mountain.


\bibliography{sirio}


\appendix

\section{Telluric Correction}
\label{appendix:telluric}

\begin{figure*}[tbp]
\includegraphics[width=\textwidth]{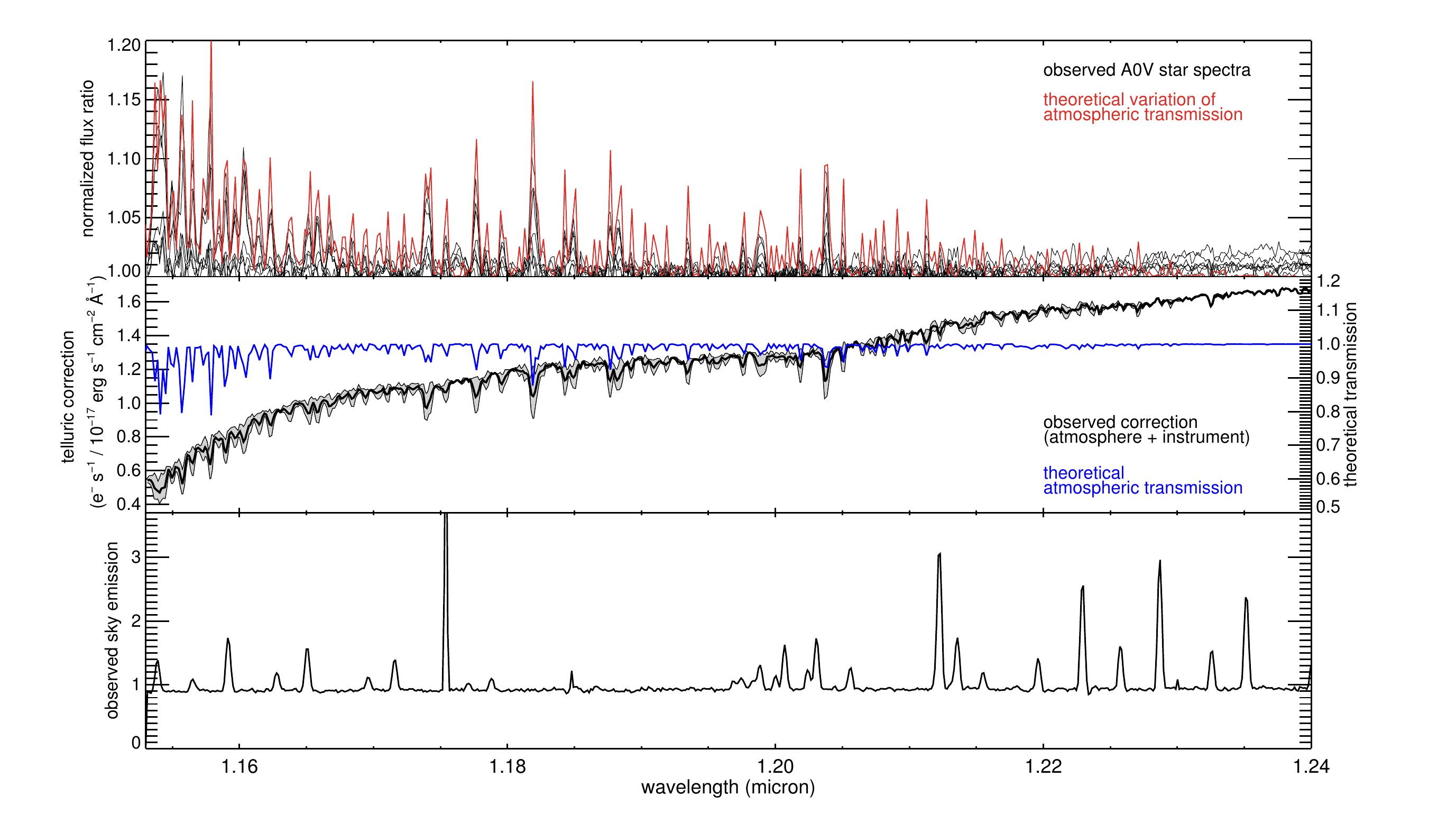}
\caption{Example of telluric correction and its uncertainty, for a spectral region in the $J$ band. These plots are based on 13 observations of two A0V stars taken in different conditions, local times, and nights. Top: observed fluxes, normalized by their median value. The discrepancy between different observations presents significant spikes, which are well described by a simple model that accounts for typical variations in airmass and water vapor (shown in red). Center: combining all the observed spectra gives a master telluric correction, shown in black. Its uncertainty. plotted in gray, is derived by adopting the theoretical curve for the variations in the atmospheric transmission. The detailed features in the telluric correction match well the theoretical curve for the transmission of the atmosphere (shown in blue). Bottom: Median sky spectrum from the stars observations, normalized to its continuum.}
\label{fig:telluric_example}
\end{figure*}

The atmosphere introduces significant asorption in the near-infrared. To correct for this absorption we observed the same two A0V stars (HIP55627 and HIP98640) throughout the three MOSFIRE runs. The standard stars were observed most of the nights, at the beginning and/or at the end of the night.

Because the atmospheric transmission depends on airmass and water vapor, we expect a variation in the telluric features for different observations. We estimate such variation by comparing the observed spectra of the standard stars taken at different nights. Since A0 stars have a very smooth spectrum, we can combine the data for the two stars, as virtually all the features observed are due to telluric absorption, with the only exception of a few Hydrogen absorption lines. For each band, we have between 8 and 14 individual spectra spanning a variety of nights and atmospheric conditions. By comparing these data we conclude that the telluric correction is generally very stable with airmass and time (both within one night and among different nights), except at the edges of the near-infrared bands, where the absorption is significantly stronger. This can potentially cause an imperfect correction if the standard star and the science targets were not observed in the same exact conditions. To assess this effect, we make use of the ATRAN models \citep{lord92} of the atmospheric transmission spectrum, which we downloaded from the Gemini website\footnote{http://www.gemini.edu}. We select the two examples that represent extreme conditions of the atmosphere: airmass 1 and 1 mm of water vapor, and airmass 2 and 3 mm of water vapor. We take half of their difference as an approximate measure of the typical variation in the telluric spectrum between the standard star and the science observations. In the top panel of Figure \ref{fig:telluric_example} we compare this theoretical variation in the telluric absorption to the observed one, for a spectral region in the $J$ band. We found that the model is an excellent prediction of the observed variation in all four near-infrared bands.

The telluric correction is calculated for each A0 star observation by following the method of \citet{vacca03} and \citet{cushing04}. A high-resolution model of Vega is broadened and shifted in order to match the intrinsic spectrum of the standard stars, including the detailed shape of the Hydrogen absorption lines. This procedure yields a very accurate template of the intrinsic spectrum of the star, which is then divided by the observed flux to obtain the telluric correction. For each band, the individual results are then combined into a master telluric correction, shown in the central panel of Figure \ref{fig:telluric_example}. This curve accounts for both the instrumental response and the atmospheric absorption.

\begin{figure*}[tbp]
\includegraphics[width=\textwidth]{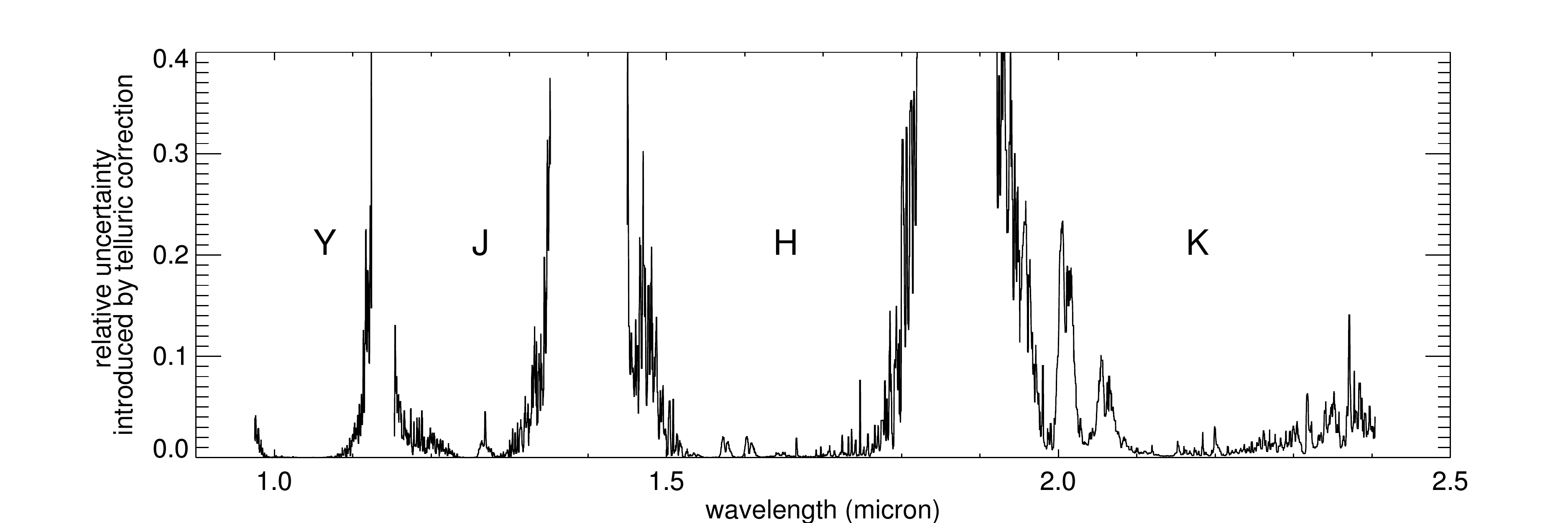}
\caption{Relative uncertainty on the calibrated flux introduced by the telluric correction, median-smoothed. Using a universal telluric correction increases the flux error by an amount that is almost always smaller than 10\%. The exceptions are near the edges of the atmospheric windows and in the CO$_2$ bands around 2$\mu$m.}
\label{fig:telluric_error}
\end{figure*}

Because the variation in the telluric absorption is not very large for most of the wavelength range, we decide to use the same master telluric correction for all the observations, instead of correcting every observation with a telluric derived in the same night. This method has the advantage of a much more robust correction. However, if an observation was taken in conditions very different from the average ones, the telluric correction will introduce an error on the calibrated flux. We explicitly account for this by adopting the theoretical variation in the atmospheric absorption (red curve in the top panel of Figure \ref{fig:telluric_example}) as relative telluric uncertainty, which we add in quadrature to the flux uncertainty output by the data reduction pipeline. Figure \ref{fig:telluric_error} shows that this procedure increases the spectral error by less than 10\% for most of the wavelength range. The exceptions are the CO$_2$ bands in $K$, and the edges of the atmospheric windows, where spectra are typically not usable due to the very low transmission.

As clearly shown in Figure \ref{fig:telluric_example}, the peaks in the telluric variation correspond to the absorption lines in the atmosphere, which are mainly due to water vapor, and do not correlate with the sky emission lines, which are caused by OH. The strong variation of the OH lines is what generally dominate the error spectrum of faint objects in the near-infrared. Interestingly, we also find that the overall shape of the telluric spectra varies throughout the different observations, as can be seen in the longer wavelength region of the top panel of Figure \ref{fig:telluric_example}. Although the origin of this small variation is not known, it does not have any impact on our analysis because when fitting each spectrum we always include a low-order polynomial correction to account for a potential continuum mismatch between the data and the templates.


\section{How Representative is the MOSFIRE Sample?}
\label{appendix:samplebias}

\begin{figure*}[tbp]
	\centering
	\begin{subfigure}
		\centering
		\includegraphics[width=0.45\textwidth]{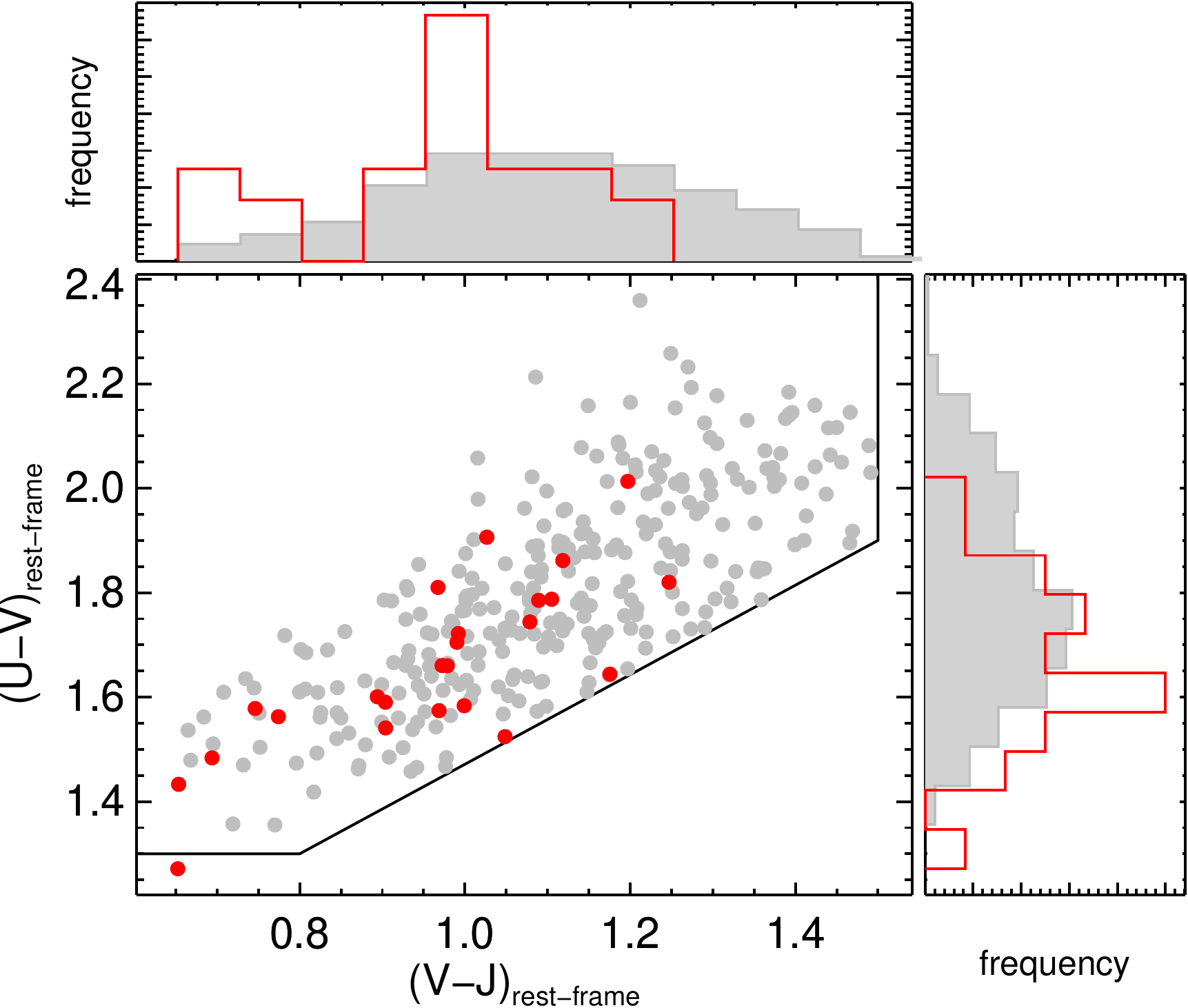}
	\end{subfigure}
	\begin{subfigure}
		\centering
		\includegraphics[width=0.45\textwidth]{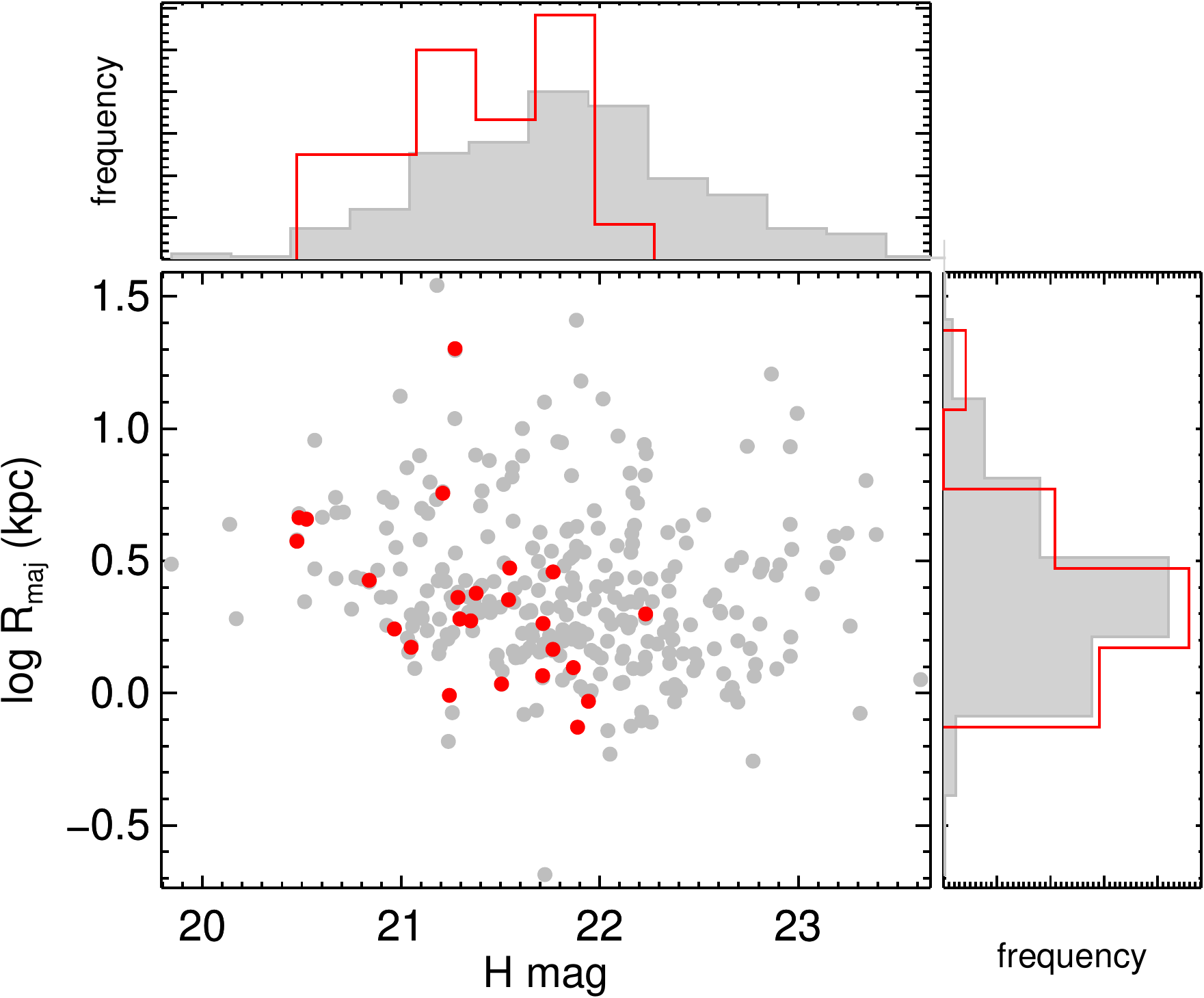}
	\end{subfigure}
	\caption{Comparison of the MOSFIRE sample (red) with a mass-complete parent population taken from the 3D-HST catalog (gray). Left: comparison of rest-frame colors on the $UVJ$ plane. Right: comparison of observed $H$ band magnitude and effective size. The right and top panels show the sample distributions collapsed onto a single axis.}
	\label{fig:samplebias}
\end{figure*}

Since spectroscopic surveys are typically biased toward brighter targets, it is important to understand how representative of the underlying population our MOSFIRE sample is. For this purpose we compare the properties of the galaxies in our sample with those of a sample extracted from the 3D-HST catalog \citep{brammer12, skelton14}, which is virtually complete for the population of massive galaxies at $z<2.5$. The lack of spectroscopic redshifts does not represent an issue, because the large number of photometric data points allow the derivation of reliable photometric redshifts. We construct a parent catalog in the following way: we select all galaxies in the $UVJ$ quiescent box as defined by \citet{muzzin13} with stellar masses above $10^{10.8} \Msun$ and redshift in the range $1.5 < z < 2.5$. This selection includes naturally almost the entire MOSFIRE sample, since the initial target selection was done on the 3D-HST catalog following similar criteria (as discussed in Section \ref{sec:data_selection}). In the left panel of Figure \ref{fig:samplebias} we compare the distribution of these two samples in the $UVJ$ diagram. The MOSFIRE sample probes most of the red sequence but misses very red galaxies, which are old and/or dusty, and therefore fainter. This is also clear from the right panel, where the distribution in brightness and size is shown. The MOSFIRE and the parent samples have different distributions of $H$ magnitude values, as confirmed by a K-S test (p=0.002): virtually all galaxies fainter than $H\sim22$ are missed by the MOSFIRE observations. However, the size distribution of the detected objects is not significantly different from that of the parent population (p=0.94). We also checked that this size comparison is not affected by the mismatch in the brightness distribution. To do this, we repeated the comparison selecting only targets brighter than $H_\mathrm{lim}=21.9$ (this excludes two galaxies from the MOSFIRE sample). The distributions of H magnitudes for the two samples are now similar (a K-S test yields p=0.80), and the size distributions are largely unchanged and still consistent with each other (p=0.56). This result is qualitatively the same if we change $H_\mathrm{lim}$ by $\pm0.1$. We conclude that the MOSFIRE sample is not biased against larger objects.

\end{document}